\documentclass[journal]{IEEEtran}

\usepackage{url}
\usepackage{amsmath,graphicx} 
\usepackage{array}
\newcolumntype{?}{!{\vrule width 1pt}} 
\usepackage{multicol}
\usepackage{pifont}
\usepackage{amssymb}
\newcommand{\cmark}{\text{\ding{51}}}
\newcommand{\xmark}{\text{\ding{55}}}
\usepackage{tabu}

\usepackage{arydshln}

\usepackage{url}
\usepackage{booktabs}
\usepackage{multicol}
\usepackage{multirow}
\usepackage{tikz,tikzscale}
\usepackage{pgfplots}
\pgfplotsset{compat=newest, width = \textwidth}
\usepgfplotslibrary{groupplots}
\usepackage{amsmath,amsfonts,amssymb,amsthm, bm}
\usepackage[ruled,vlined,linesnumbered ]{algorithm2e}
\usepackage{ctable}

\usepackage{cite}
 \ifCLASSINFOpdf
\else
\fi

\usepackage{algorithmic}
\usepackage{soul}
\usepackage{color}

\begin{document}
\title{Energy Efficient Video Decoding for VVC Using a Greedy Strategy Based Design Space Exploration}%

\author{Matthias~Kr\"anzler, 
Christian Herglotz,~\IEEEmembership{Member,~IEEE,}
 and~Andr\'e~Kaup,~\IEEEmembership{Fellow,~IEEE}
 \thanks{Manuscript received August 27, 2021; revised November 10, 2021; accepted November 20, 2021. Corresponding author: Matthias~Kr\"anzler.}
\thanks{The authors are with the Chair of Multimedia Communications and Signal Processing, Friedrich-Alexander-Universit\"at Erlangen-N\"urnberg (FAU), Erlangen, Germany (e-mail: matthias.kraenzler@fau.de; christian.herglotz@fau.de; andre.kaup@fau.de). Digital Object Identifier 10.1109/TCSVT.2021.3130739.}}

\markboth{IEEE Transactions on Circuits and Systems for Video Technology, November~2021}%
{Journal}

\newcommand{\pro}[1]{\mbox{#1\,\%}}
\newcommand{\vscale}[1]{}
\definecolor{amaranth}{rgb}{0.9, 0.17, 0.31}

\maketitle

\IEEEpubid{\begin{minipage}{\textwidth}\ \\ \\ \\ \\[8pt] \centering
Copyright $\copyright$ 20xx IEEE. Personal use of this material is permitted. However, permission to use this material for any other purposes must be obtained from the IEEE by sending an email to pubs-permissions@ieee.org.
\end{minipage}}

\begin{abstract}
IP traffic has increased significantly in recent years, and it is expected that this progress will continue. Recent studies report that the viewing of online video content accounts for a share of \pro{1} of the global greenhouse gas emissions. To reduce the data traffic of video streaming, the new standard Versatile Video Coding (VVC) has been finalized in 2020. In this paper, the energy efficiency of two different VVC decoders is analyzed in detail. Furthermore, we propose a design space exploration that uses an algorithm based on a greedy strategy to derive coding tool profiles that optimize the energy demand of the decoder. We show that the algorithm derives optimal coding tool profiles for a subset of coding tools. Additionally, we propose profiles that reduce the energy demand of VVC decoders and provide energy savings of more than \pro{50} for sequences with 4K resolution. Thereby, we will also show that the proposed profiles can have a lower decoding energy demand than comparable HEVC-encoded bit streams while also having a significantly lower bit rate.
\end{abstract}


\IEEEpeerreviewmaketitle

\section{Introduction}
\IEEEPARstart{I}{n} recent years, IP traffic has increased significantly, and by 2022, it will rise by another \pro{60} in relation to 2020~\cite{CSI2019}. A key factor in this development is the increased usage of internet video streaming, which will have a traffic share of over \pro{80} in 2022. The gained popularity of video-on-demand (VoD) services and the increased use of video conferencing resulting from the COVID pandemic have an amplifying effect on the IP traffic. In addition, the growth in video streaming demand is driven by a larger proportion of video devices with higher resolution such as Ultra High Definition (UHD) TVs and the rising speed of broadband and mobile connections~\cite{CSI2019}.

Simultaneously, it is reported that the viewing of online video content had a share of \pro{1} of the global greenhouse gas (GHG) emissions in 2018~\cite{Efoui-Hess2019}, which is comparable to the total emissions of a country such as Spain. Considering the previously mentioned increase in video streaming, it can be assumed that this share will increase as well. Consequently, it is essential to improve the energy efficiency of video coding systems in order to reduce GHG emissions. Energy efficiency is not only necessary from an environmental perspective, but also for battery powered mobile devices, such as smartphones, which have a limited battery lifetime~\cite{Herglotz19a}.

To reduce the global IP traffic caused by video communication and to manage the increased demand for immersive video technology such as Virtual Reality~(VR), the Versatile Video Coding~(VVC) standard was finalized in 2020. The goal of VVC is to reduce the bit rate by approximately \pro{50} at an equal subjective visual quality compared to its predecessor High Efficiency Video Coding~(HEVC)~\cite{Bross2021a}. The improvement of the compression performance in terms of rate-distortion~(RD) efficiency results in a significant rise in computational complexity and energy demand of both the encoder and the decoder. In previous works, it is reported that the energy demand of the reference decoder implementation of VVC is increased by over \pro{80} relative to HEVC~\cite{Kraenzler2020MMSP}. Therefore, we propose a solution to improve the energy efficiency of VVC that has a higher energy efficiency than HEVC.

The energy demand of mobile devices for video streaming applications is modeled in\mbox{\cite{HerglotzCoulombeVazquezEtAl2020}}. It is shown that the display brightness, the frame rate, and the streaming bit rate are well suited to model the energy demand accurately. The authors conclude that those components have a major influence on the energy efficiency of video streaming. In~\cite{KhernacheBenmoussaBoukhobzaEtAl2021}, the authors analyze the energy demand of HEVC software and hardware decoders on different devices. It turns out that hardware decoders consume up to~\pro{30} of the total device energy and for software decoding, this share is up to~\pro{50}. However, the authors found that the energy demand highly depends on the resolution and frame rate of the video sequence, similar to the findings in~\cite{HerglotzCoulombeVazquezEtAl2020}. Therefore, with a higher resolution or frame rate, the ratio of hardware and software decoding power increases. For a resolution such as 1080p, which is common for mobile devices, the ratio remains below four. According to the authors, the ratio between hardware and software decoding can be further reduced, when the software decoder gets more efficient, which is one aspect that we propose in this paper.

In literature, one can find several approaches that reduce the energy demand of a video decoder. In~\cite{NoguesMenardPelcat2019}, the authors propose an algorithmic method for HEVC, which utilizes two modifications towards the regular HEVC decoder to reduce the energy demand by up to \pro{40}. First, the complexity of the motion compensation filters is simplified by reducing the length of the filters. Second, in-loop and motion compensation filters are skipped with a specific ratio. Another approach that addresses the motion compensation and deblocking filter operations is evaluated in~\cite{Yang2018}. The authors propose a method that is capable of reducing the decoder complexity at a specified rate. Furthermore, to improve the subjective quality of the decoded images, those actions are mainly applied to non-salient areas, which are derived with the approach presented in~\cite{Yang2016}. Thereby, the authors achieve complexity reductions from~\pro{15} up to~\pro{40}.

Alternatively, the decoder's energy demand can be reduced by lowering the processor's clock frequency while keeping the frame rate of the output video. This method is called Dynamic Voltage and Frequency Scaling (DVFS)~\cite{NoguesRaffinPelcatEtAl2015}. With a lower processing frequency, the energy demand is reduced without a loss in visual quality.

Despite the previous methods, not only the decoder itself can be optimized, but also the encoder can be utilized to optimize the energy demand of the subsequent decoding process. This is achieved by changing the rate-distortion optimization (RDO) in the encoder. Therefore, RDO has to be adjusted to the energy domain~\cite{MallikarachchiTalagalaH.EtAl2017,CorreaCorreaPalominoEtAl2018,HerglotzHeindelKaup}. 

In~\cite{MallikarachchiTalagalaH.EtAl2017}, the complexity, which is closely related to the processing energy, is modeled in terms of CPU cycles that are needed for decoding. With this optimization, it is possible to reduce the energy demand of the decoder by \pro{41}, while the peak signal-to-noise ratio (PSNR) decreases by approximately 2.5~dB.

In~\cite{CorreaCorreaPalominoEtAl2018}, the energy is modeled in terms of the time demand to execute the decoder. Thereby, the authors report an average energy reduction of approximately \pro{10}, a bit rate increase of \pro{10}, and a PSNR reduction of 0.26~dB.

Alternatively, the energy demand of the decoder is estimated in advance in the encoder with a bit stream feature-based energy model~\cite{HerglotzSpringerReichenbachEtAl2018}, which is called Decoding-Energy-Rate-Distortion-Optimization (DERDO)~\cite{HerglotzHeindelKaup}. The authors achieve an energy reduction of up to \pro{30} for a maximum bit rate increase of \pro{50}. Therefore, for each video codec, a bit stream model is necessary. In~\cite{Kraenzler2019} and~\cite{Kraenzler2020ICIP}, corresponding bit stream models are proposed for HEVC and VVC, respectively.

In literature, several energy models describe the energy demand of a smartphone while decoding video content~\cite{Li12} or virtual reality content~\cite{Herglotz19b}. These models can also be utilized to optimize energy consumption.

In addition to the decoder, the encoder complexity is optimized in\mbox{\cite{Li2011,Huang2020a,YangShenDongEtAl2020}}. The classical RDO is extended by Li \textit{et~al.}\mbox{\cite{Li2011}} with a metric that describes the complexity of the coding mode testing in Advanced Video Coding (AVC), through the rate-complexity-distortion optimization (RCDO). With RCDO it is possible to adaptively control encoder complexity. For HEVC, Huang \textit{et~al.}\mbox{\cite{Huang2020a}} propose an improved RCDO algorithm that utilizes a constrained mathematical description of the performance to trade-off complexity and compression efficiency. Furthermore, several approaches are applied that improve the coding mode decision process.

In\mbox{\cite{YangShenDongEtAl2020}}, an algorithm is proposed that can reduce the complexity of the intra mode by an early skipping of intra prediction and partition modes in VVC. Furthermore, a fast intra mode decision is proposed that is based on a gradient descent search, which is a greedy algorithm. 

The block partitioning that is used by VVC is presented in\mbox{\cite{HuangAnHuangEtAl2021}}. The authors show that the usage of the binary tree and ternary tree partitioning has a huge influence on the compression efficiency and encoding time (over 6-times compared to HEVC), but little influence on the decoding time, which increases in the range of 2-\mbox{\pro{4}}. Hence, we neglect the partitioning in this work.

For this paper, we develop an approach based on the findings in~\cite{Kraenzler2020MMSP}, where it was determined that several coding tools increase the compression efficiency by adding computational complexity and thereby, increase the energy demand of the decoder as well. However, it was also discovered that some coding tools even reduce the energy demand. In this paper, the evaluation of coding tools in terms of their energy efficiency will be studied with a tool sensitivity analysis showing both the impact on rate-distortion efficiency and energy efficiency.  

Furthermore, we propose to use a novel design space exploration (DSE) to optimize the energy demand. The DSE will derive coding tool profiles, which indicate the usage of each considered coding tool.

In summary, this work will provide the following contributions:
\begin{itemize}
	\item General energy efficiency analysis of several decoder implementations of VVC and HEVC.
	\item Methodology to derive optimum energy efficient coding tool profiles.
	\item Analysis of greedy strategy based algorithm on a subset of features
	\item Tool sensitivity analysis of VTM decoder.
	\item Energy efficient profiles for VVC with superior energy efficiency to HEVC.
\end{itemize}

In Section~\ref{sec:codingTools}, we will give a brief overview of VVC and show for each coding configuration (RA, LB, AI), which coding tools are enabled. Afterward, Section~\ref{sec:DSE} presents the proposed design space exploration for the determination of energy efficient coding tool profiles. Then, in Section~\ref{sec:metrics}, we will explain the energy measurement setup, the evaluation metrics, the used test sequences, and the software setup with the used video codec implementations of encoder and decoder. Thereafter, we will analyze the energy efficiency of two VVC decoders, namely VTM and VVdeC~\cite{VVdec}, in detail and compare the results to comparable HEVC decoders. Furthermore, in Section~\ref{sec:evaluation}, we will show the results of the proposed algorithm in comparison to a full search method for subset of coding tools. Then, we will evaluate the DSE in detail and propose two coding tool profiles for each coding configuration. Finally, Section~\ref{sec:conclusion} will conclude this paper.

\section{Overview on VVC and its Coding Tools}
\label{sec:codingTools}

\begin{table}[t!]
\def\arraystretch{1.01}
\caption{All considered coding tools of VVC are divided into five groups, which are Intra, Inter, Transformation/Quantization, In-Loop Filter, and Others. All tools that are initially enabled for a specific coding tool profile are denoted by~$\cmark$, all tools that are disabled by the encoder by~$\xmark$, and tools that cannot be used by the specific profile \mbox{by -}. In this paper, we use the coding configurations all intra~(AI), randomaccess~(RA), and lowdelay~B~(LB). $\eta$ corresponds to the index of a coding tool.}

\label{tab:ToolList}
\begin{center}
\begin{tabular}{@{}cl|ccc@{}} 
 $\eta$ & \multicolumn{1}{c|}{Tool}    & AI& LB&RA\,\\
\hline\hline
\multicolumn{5}{c}{Intra}  \\
\hline\hline
1 &  Cross-component linear model (CCLM)         &$\cmark$&$\cmark$&$\cmark$\\ 
2 &  Intra sub-partition (ISP)                   &$\cmark$&$\cmark$&$\cmark$\\ 
3 &  Matrix-based intra-picture prediction (MIP) &$\cmark$&$\xmark$&$\cmark$\\ 
4 &  Multiple reference line (MRL)               &$\cmark$&$\cmark$&$\cmark$\\ 
 \hline\hline
\multicolumn{5}{c}{Inter} \\
\hline\hline
5 & Affine motion (AFFINE)                      &-&$\cmark$&$\cmark$\\
6 & Adaptive MV resolution (AMVR)               &-&$\cmark$&$\cmark$\\
7 & Biprediction with CU-level weights (BCW)    &-&$\cmark$&$\cmark$\\
8 & Bidirectional optical flow (BDOF)   &-&-&$\cmark$\\
9 & Combined inter-/intra-picture prediction (CIIP) &-&$\cmark$&$\cmark$\\
10 & Decoder-side MV refinement (DMVR)              &-&-&$\cmark$\\
11 & Geometric partitioning mode (GPM)              &-&$\cmark$&$\cmark$\\
12 & Merge with MVD (MMVD)                          &-&$\cmark$&$\cmark$\\
13 & Prediction refinement with optical flow (PROF) &-&$\cmark$&$\cmark$\\
14 & Subblock-based temporal MVP (SBTMVP) &-&$\cmark$&$\cmark$\\
15 & Symmetric MVD (SMVD)   &-&-&$\cmark$\\
\hline\hline
\multicolumn{5}{c}{Transformation and Quantization} \\
\hline\hline
16 & Dependent quantization (DQ) &$\cmark$&$\cmark$&$\cmark$\\
17 & Joint coding of chroma residual (JCCR) &$\cmark$&$\cmark$&$\cmark$\\
18 & Low-frequency non-separable transform (LFNST)  &$\cmark$&$\xmark$&$\cmark$\\
19 & Multiple transform selection (MTS)    &$\cmark$&$\cmark$&$\cmark$\\
20 & Subblock transform (SBT)    &-&$\cmark$&$\cmark$\\
\hline\hline 
\multicolumn{5}{c}{In-Loop Filter} \\
\hline\hline
21 & Adaptive loop filter (ALF)    &$\cmark$&$\cmark$&$\cmark$\\
22 & Cross-component adaptive loop filter (CCALF)  &$\cmark$&$\cmark$&$\cmark$\\
23 & Deblocking filter (DBF)    &$\cmark$&$\cmark$&$\cmark$\\
24 &  Luma mapping with chroma scaling (LMCS)   &$\cmark$&$\cmark$&$\cmark$\\
25 &  Sample adaptive offset (SAO) &$\cmark$&$\cmark$&$\cmark$\\
\hline\hline
\multicolumn{5}{c}{Others} \\
\hline\hline
26 & Block-level differential PCM (BDPCM)  &$\xmark$&$\xmark$&$\xmark$\\
27 & Intra-picture block copy (IBC)    &$\xmark$&$\xmark$&$\xmark$\\
28 & Chroma separate tree (CST)    &$\cmark$&$\cmark$&$\cmark$\\
 \hline \hline
\end{tabular}
\end{center}
\end{table}

In this section, we will give a short overview of all coding tools that we used for our energy analysis. As mentioned before, the goal of VVC is to reduce the bit rate in relation to its predecessor HEVC by \pro{50}. This reduction is mainly achieved by various coding tools and a changed partitioning scheme. The partitioning of coding tree units~(CTUs) in VVC is realized by a quadtree structure, which was previously used in HEVC~\cite{Sullivan2012}, and an additional nested multi-type tree~(MTT)~\cite{JVET-Q2002} splitting, which allows splitting into non-square blocks. 

Besides the enhanced partitioning scheme, VVC introduces a significant number of coding tools. All coding tools that were considered for the evaluation in this paper are shown in Table~\ref{tab:ToolList}. In this table, it is denoted whether a coding tool is enabled according to the CTC configuration of the VTM encoder~\cite{VTM}. Throughout the paper, the usage settings of Table~\ref{tab:ToolList} will be referred to as CTC profile. Tools that are initially enabled are denoted by a ($\cmark$) and disabled tools are denoted by \mbox{a ($\xmark$)} in Table~\ref{tab:ToolList}. Furthermore, coding tools that are not relevant for the specific coding tool profile are denoted \mbox{by (-)} (e.g., inter prediction tools for AI). A detailed description of the corresponding coding tools can be found in~\cite{Bross2021}.

\section{Design Space Exploration for the Derivation of Energy Efficient Profiles}
\label{sec:DSE}

In the following, we will describe the basic concept to achieve increased energy efficiency with a design space exploration (DSE)~\cite{Rosales_2016}. In~\cite{Rosales_2016}, the encoder's complexity of the HEVC mode decision process is optimized by parallelization and skip decisions. In contrast to~\cite{Rosales_2016}, we will focus on coding tools and the energy demand reduction of the decoder. Based on the findings in~\cite{Kraenzler2020MMSP}, we disable and enable coding tools of VVC in the encoder and thereby, reduce the decoding energy demand.

We define the design space by the tradeoff between energy and compression efficiency. To achieve a higher energy efficiency for VVC, we introduce the coding tool profile~$\boldsymbol{u}$, which is defined by
\begin{equation}
\boldsymbol{u} = \left( \begin{array}{c} u(1) \\\ \vdots  \\ u(\eta) \\ \vdots  \\\ u(N)  \end{array}\right), 
\end{equation}
where~$u(\eta)$ indicates the usage of a coding tool, $\eta$ corresponds to the index of a specific coding tool, and $N$ to the number of coding tools from Table~\ref{tab:ToolList}. Each entry represents a binary value~$u(\eta)\in \left\lbrace0,1\right\rbrace$ indicating whether the tool is disabled or enabled. For the initialization of $\boldsymbol{u}$, we consider 28 coding tools according to Table~\ref{tab:ToolList}. For each coding tool~$\eta$ that is marked with $(\cmark)$, $u(\eta)$ is 1. Otherwise, for a tools that is marked with $(\xmark)$, $u(\eta)$ is $0$, and the remaining tools marked with~$(-)$ are not considered.

To evaluate the influence of a changed coding tool profile on energy and compression efficiency, we use the Bj{\o}ntegaard-Delta~(BD) metric. With a BD-metric, we compare the efficiency of an arbitrary video codec to another codec. Each BD metric that we use in this paper is based on the Bj{\o}ntegaard-Delta~bit~rate~(BDR-PSNR)~\cite{VCEG-M33}, which describes the bit rate savings in \% for the same objective quality measured in PSNR. To evaluate the energy efficiency of a decoder, we substitute the bit rate by the decoder's energy demand. We call the resulting BD metric Bj{\o}ntegaard-Delta~decoding~energy~(BDDE-PSNR), which describes the energy savings in \% for the same PSNR. The measurement of the decoding energy and of PSNR will be explained in detail in Section~\ref{sec:metrics}.

 For the optimization of the decoding energy demand, we iteratively change the coding tool profile~$\boldsymbol{u}$, encode the sequences with four different QPs, and subsequently measure the decoding energy of the resulting bit streams. Mathematically, we model this process using the function~$f(\boldsymbol{u})$ with $\boldsymbol{u}$ as input with
\begin{equation}
\min_{\boldsymbol{u}}~\textrm{BDDE-PSNR}  = \min_{\boldsymbol{u}}~f\left(\boldsymbol{u} \right),
\label{eq:min}
\end{equation}
where BDDE-PSNR is calculated using the CTC coding tool profile from Table~\ref{tab:ToolList} as a reference.

To find a solution for~\eqref{eq:min}, a full search is a possible solution. However, in this case, it would be necessary to encode and measure the decoding energy demand of all combinations, which has a complexity of~$\mathcal{O}\left(2^n\right) = 2^{28} =  268{,}435{,}456$ coding tool profiles for RA coding, which is not feasible in practice. 

\begin{algorithm}[!t]
\begin{small}
\SetAlgoLined
 Initialize $\boldsymbol{u}_{\boldsymbol{1,0}}$ \;
 $i$ = 1 \;
\While{$\boldsymbol{u}_{\boldsymbol{i,0}} \neq \boldsymbol{u}_{\boldsymbol{i+1,0}}$}{

    ${\text{PSNR}_{i,0},\text{Energy}_{i,0}} \gets \text{Analyze}\left(\boldsymbol{u}_{\boldsymbol{i,0}}\right)$\;
 \For{$\nu = 1,2,...,N$}{
  $\boldsymbol{u}_{\boldsymbol{i,\nu}} = \boldsymbol{u}_{\boldsymbol{i,0}}$\;
   \eIf{$\boldsymbol{u}_{\boldsymbol{i,0}}\left( \nu \right) ==  0$}
   {
     $\boldsymbol{u}_{\boldsymbol{i,\nu}}\left( \nu \right) = 1 $\; 
   }{ 
     $\boldsymbol{u}_{\boldsymbol{i,\nu}}\left( \nu \right) = 0 $\;
   }

    ${\mathrm{PSNR}_{i,\nu},\text{Energy}_{i,\nu}} \gets \text{Analyze}\left( \boldsymbol{u}_{\boldsymbol{i,\nu}}\right)$\;  
   Calculate $\mathrm{BDDE}\left(\boldsymbol{u}_{\boldsymbol{1,0}} , \boldsymbol{u}_{\boldsymbol{i,\nu}}\right)$\;
    \If{$\mathrm{BDDE}\left( \boldsymbol{u}_{\boldsymbol{1,0}} , \boldsymbol{u}_{\boldsymbol{i,\nu}} \right) < \mathrm{BDDE}\left(\boldsymbol{u}_{\boldsymbol{1,0}} ,\boldsymbol{u}_{\boldsymbol{i,0}}\right)$}
   {
        $\boldsymbol{u}_{\boldsymbol{i+1,0}}\left( \nu \right) = \boldsymbol{u}_{\boldsymbol{i,\nu}} \left( \nu \right)$\;
   }
    }
 \If{$ \mathrm{BDDE} \left(\boldsymbol{u}_{\boldsymbol{1,0}} , \boldsymbol{u}_{\boldsymbol{i,\nu}}\right) \geq \mathrm{BDDE}\left(\boldsymbol{u}_{\boldsymbol{1,0}} ,\boldsymbol{u}_{\boldsymbol{i-1,0}} \right) \forall  \nu$} 
 { 
  Break\;
 }
    $i = i + 1$ \;
}
\end{small}
 \caption{DSE Optimization Algorithm} 
 \label{alg:Gradient}
\end{algorithm}

Therefore, we propose a novel approach that is a greedy strategy based iterative Design Space Iteration (DSE) to determine coding tool profiles that increase the energy efficiency. Thereby, we can reduce the complexity to determine coding tool profiles to $\mathcal{O}\left(i \cdot n\right)$, where $i$ corresponds to the number of iterations. Consequently, the complexity can be reduced significantly. However, as the algorithm does not guarantee optimality, we will evaluate the algorithm in detail for optimality in Subsection~\ref{subsec:DSE_Eval} using a subset of tools. The encoded bit streams of the proposed coding tool profiles can be decoded with any decoder that is conforming with the specification of VVC.

In Algorithm~\ref{alg:Gradient}, the proposed method is described. We define the vector~$\boldsymbol{u}_{\boldsymbol{i,\nu}}$, where $i$ corresponds to an iteration of the while loop (c.f. line~3 in Algorithm~\ref{alg:Gradient}) and $\nu$ to a coding tool within the for loop (c.f. line~5 in Algorithm~\ref{alg:Gradient}). If $\nu$ equals zero, $\boldsymbol{u}_{\boldsymbol{i,0}}$ corresponds to the reference of an iteration~$i$. For the calculation of BDDE in Algorithm~\ref{alg:Gradient}, we use the CTC profile ($\boldsymbol{u}_{\boldsymbol{1,0}}$) as a reference. Furthermore, the algorithm gets the values of PSNR and decoding energy demand from the function Analyze$\left( \boldsymbol{u}_{\boldsymbol{i,0}}\right)$ (c.f. line~4 and 12 in Algorithm~\ref{alg:Gradient}).

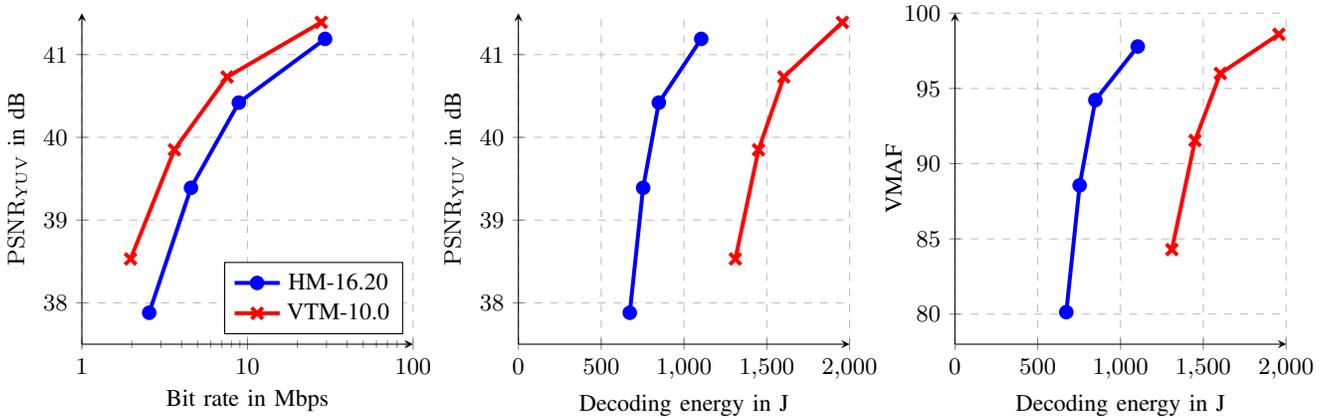
\begin{figure*}[t!]
\begin{small}
\begin{center}
\begin{tikzpicture}
\begin{groupplot}[
     group style = {group size = 3 by 1},
     height = 10cm,
     width = \textwidth,
    ]
\nextgroupplot[
    xmode=log,
	width=0.33\textwidth,
	height = 0.33\textwidth,
    log ticks with fixed point,
    xlabel={Bit rate in Mbps},
    ylabel={$\mathrm{PSNR}_{\mathrm{YUV}}$ in dB},
    xmin=1, xmax=100,
    ymin=37.5, ymax=41.5,
    legend pos=south east,
    ymajorgrids=true,
    xmajorgrids=true,
    grid style=dashed,
    axis lines = left,
]
  \hspace*{-0.4cm}
\addplot[
    color=blue,
    mark=*,
    line width=1.5pt,
    ]
    coordinates {
    (2.55,37.88)(4.56,39.39)(8.87,40.42)(29.42,41.19)
    };
    \addlegendentry{HM-16.20}

\addplot[
    color=red,
    mark=x,
    mark options={scale=1.5, fill=red},
    line width=1.5pt,
    ]
    coordinates {
    (1.968,38.53)(3.620,39.85)(7.536,40.73)(27.857,41.39)
    };
    \addlegendentry{VTM-10.0}

\nextgroupplot[
	width=0.33\textwidth,
	height = 0.33\textwidth,
    xlabel={Decoding energy in J},
    ylabel={$\mathrm{PSNR}_{\mathrm{YUV}}$ in dB},
    xmin=0, xmax=2000,
    ymin=37.5, ymax=41.5,
    legend pos=north west,
    ymajorgrids=true,
    xmajorgrids=true,
    grid style=dashed,
    axis lines = left,
]
  \hspace*{0.4cm}
\addplot[
    color=blue,
    mark=*,
    line width=1.5pt,
    ]
    coordinates {
    (672.64,37.88)(753.15,39.39)(848.09,40.42)(1103.9,41.19)
    };

\addplot[
    color=red,
    mark=x,
    mark options={scale=1.5, fill=red},
    line width=1.5pt,
    ]
    coordinates {
    (1310,38.53)(1449.9,39.85)(1603.0,40.73)(1956.4,41.39)
    };

\nextgroupplot[
	width=0.33\textwidth,
	height = 0.33\textwidth,
    xlabel={Decoding energy in J},
    ylabel={VMAF},
    xmin=0, xmax=2000,
    ymin=78, ymax=100,
    ymajorgrids=true,
    xmajorgrids=true,
    grid style=dashed,
    axis lines = left,
    y label style={at={(-0.134,0.5)}}
]
 \hspace*{0.4cm}
\addplot[
    color=blue,
    mark=*,
    line width=1.5pt,
    ]
    coordinates {
    (672.64,80.13)(753.15,88.56)(848.09,94.23)(1103.9,97.79)
    };

\addplot[
    color=red,
    mark=x,
    mark options={scale=1.5, fill=red},
    line width=1.5pt,
    ]
    coordinates {
    (1310,84.29)(1449.9,91.54)(1603.0,96.00)(1956.4,98.60)
    };
 
\end{groupplot}

\end{tikzpicture}
\end{center}
\end{small}
\vspace*{-0.5cm}

\caption{Evaluation of the sequence Tango2, which is encoded with the RA configuration, with the described metrics of Section \ref{subsec:EvalMetrics}. The blue curve corresponds to the bit streams decoded by HM-16.20 and the red curve to the bit streams decoded by VTM-10.0. The markers of each curve correspond to the QP values 22, 27, 32, and 37.}
\label{fig:BDDE}
\vspace{-0.4cm}
\end{figure*}

At first, we initialize $\boldsymbol{u}_{\boldsymbol{1,0}}$ with the values from Table~\ref{tab:ToolList} for the corresponding coding tool profile. After the bit streams are generated, we measure the decoding energy demand and analyze the bit streams in terms of PSNR. Then, for each coding tool~$\nu$ in the for loop, we change the usage of~$\boldsymbol{u}_{\boldsymbol{i,\nu}}\left(\nu\right)$. Then, we encode and evaluate the quality and the decoding energy of the corresponding bit streams. Afterward, we calculate the BDDE value, which evaluates the changes in terms of the energy demand if the coding tool is disabled or enabled. If the energy demand is decreased in relation to the reference coding tool profile of the iteration ($\mathrm{BDDE}\left(\boldsymbol{u}_{\boldsymbol{1,0}} ,\boldsymbol{u}_{\boldsymbol{i,\nu}}\right) < \mathrm{BDDE}\left(\boldsymbol{u}_{\boldsymbol{1,0}} ,\boldsymbol{u}_{\boldsymbol{i,0}}\right)$), c.f. line~14 in Algorithm~\ref{alg:Gradient}, we conclude that the reference coding tool profile of the iteration~$\boldsymbol{u}_{\boldsymbol{i,0}}$ has a lower energy efficiency and we keep the change of the~$\nu$-th coding tool for the next iteration (c.f. line~15 in Algorithm~\ref{alg:Gradient}). 

Finally, after all coding tools are evaluated for an iteration~$i$, we compare $\boldsymbol{u}_{\boldsymbol{i,\nu}}$ with $\boldsymbol{u}_{\boldsymbol{i+1,\nu}}$. If both are equal (c.f. line~3 in Algorithm~\ref{alg:Gradient}), we stop the algorithm because the algorithm has converged. Another stopping criterion is the condition in line~18, which checks whether the reference profile of the previous iteration has a higher energy efficiency than all tests of the current iteration. In Section~\ref{sec:evaluation}, we will evaluate the DSE algorithm to show that the method can achieve significant energy savings. The used software decoder should serve as an example to verify the methodology, which is not limited to a specific decoder implementation.

\section{Metrics and Test Setup}
\label{sec:metrics}

As this paper evaluates the energy performance of VVC decoding, we will explain the energy measurement setup in Subsection~\ref{subsec:Energy}. Subsequently, in Subsection~\ref{subsec:softwareSetup}, we describe our software setup. Then, in Subsection~\ref{subsec:testset}, we will show the sets of video sequences used for the evaluation of the energy efficiency of the decoders. Afterward, in Subsection~\ref{subsec:EvalMetrics}, we will focus on the quality metrics VMAF and PSNR. In Subsection~\ref{subsec:Bjontegaard}, we will describe several Bj{\o}ntegaard-Delta metrics that are based on the commonly used metric Bj{\o}ntegaard-Delta~bit~rate~(BDR)~\cite{VCEG-M33}. Finally, in Subsection~\ref{subsec:HEVCvsVVC}, we will compare the RD and energy efficiency of VVC with the VTM and VVdeC decoder in relation to HEVC with the HM and openHEVC decoder, respectively.

\subsection{Energy Measurement Setup}
\label{subsec:Energy}
For our measurements, we use a desktop PC with an \mbox{Intel i7-8700} CPU, which is a processor with x86 architecture that has six cores with a base frequency of 3.20 GHz, and CentOS 7 as an operating system (OS). For the energy measurement, we use the integrated power meter Running Average Power Limit~(RAPL)~\cite{DavidGorbatovHanebutteEtAl2010} that is built-in our CPU. 

As a second measurement setup, we use a Raspberry Pi that has a Cortex-A74 quad-core CPU, which is based on the ARM architecture, and as OS we use Raspbian. The energy demand of the device is measured with the external power meter LMG95 by ZES Zimmer. The results of the evaluation with this device will be shown in Section~\ref{subsub:vvdec}.

Similar to~\cite{HerglotzSpringerReichenbachEtAl2018}, we perform multiple measurements and verify the statistical correctness with a confidence interval test. Thereby, we ensure that our measurements are not affected by noise, which can be caused, e.g., by background processes of the OS. The statistical test is defined by
\begin{equation}
2 \cdot \frac{\sigma}{\sqrt{m}} \cdot t_{\alpha}  \left( m - 1 \right)
< \beta \cdot \overline{E_{\mathrm{dec}}}~,
\label{eq:confidence}
\end{equation}
where $m$ is the number of measurements, $\sigma$ corresponds to the standard deviation of the measurement series, $\beta$ to the maximum deviation of the energy, $\alpha$ to the probability that the condition of $\beta$ is fulfilled, $t_{\alpha}$ to the critical t-value of the Student's t-distribution, and $\overline{E_{\mathrm{dec}}}$ is the average decoding energy demand of the measurement series. We define $\beta$ to be 0.02 and $\alpha$ to 0.99, which means that with a probability of \pro{99} we have maximum deviation of \pro{2} for $\overline{E_{\mathrm{dec}}}$ from the actual energy demand. Each measurement of the active decoding energy demand~$E_{\mathrm{dec}}$ is based on two separate measurements. First, the power demand is measured during the decoding process. Afterward, the power demand is measured in idle mode for the same duration as for the decoding process. Finally, $E_{\mathrm{dec}}$ is derived by subtracting the idle energy from the decoding energy.

\subsection{Software setup}
\label{subsec:softwareSetup}

For HEVC, we use the reference software implementation HM-16.20~\cite{HM1620} for the encoding and decoding of video sequences. Additionally, we use openHEVC with the version 2.0~\cite{openHEVC} as an optimized decoder implementation.

For VVC, we use the reference software implementations VTM-8.0 and VTM-10.0~\cite{VTM} for the encoding and decoding of bit streams. Furthermore, we use an optimized software decoder implementation of VVC called Versatile Video Decoder (VVdeC)~\cite{VVdec}, which is proposed in~\cite{WieckowskiHegeBartnikEtAl2020} and targets real-time decoding. For VVdeC, we use the version 1.1.2 for the energy measurements. For the reference software implementations HM and VTM, the decoder is executed on a single core and the optimized decoders openHEVC and VVdeC use the maximum number of available threads and cores.

\subsection{Test Sequences and Encoder Configurations}
\label{subsec:testset}

For our test sequence sets, we use two sets of video sequences from the literature. First, the set from the common test conditions (CTC) of JVET~\cite{JVET-N1010} and second, the test set of the ultra video group (UVG)~\cite{MercatViitanenVanne2020}. The JVET set includes 26 sequences from $416\times240$ to 4K resolution and frame rates from 20 to 60 frames per second (fps). We divide the sequences of the UVG set into three classes. For class UVG1, the sequences have a 4K resolution and a frame rate of 120 fps. For the class UVG2, the sequences also have a 4K resolution and a frame rate of 50 fps. For class UVG3, the sequences have a full HD resolution and a frame rate of 120 fps.

The encoding configurations from the CTC of JVET are used for the encoding of the bit streams. For each sequence set, we code three coding configurations, which are All Intra (AI), Lowdelay B (LB), and Randomaccess (RA). All sequences are coded with a quantization parameter (QP) of 22, 27, 32, and 37.  For sequences with 4K resolution (A1, A2, UVG1, and UVG2), LB is not coded, and for class E, RA is not coded as recommended by~\cite{JVET-N1010}. Furthermore, we use an internal coding bit depth of 10 bit for both HEVC and VVC, and the temporal subsampling of AI is set to one, which allows to measure the decoding energy demand of all frames.

\subsection{Quality Metrics}
\label{subsec:EvalMetrics}

To evaluate the visual quality, the peak signal-to-noise ratio (PSNR) is used, which is commonly used in video coding and measures the objective visual quality~\cite{WorkingPractices}. Since we do not measure $E_{\mathrm{dec}}$ of a single color component, we use $\mathrm{PSNR}_{\mathrm{YUV}}$~\cite{WorkingPractices}, which is the weighted average of all color components as shown by the following equation

\begin{equation}
\mathrm{PSNR}_{\mathrm{YUV}} = \frac{1}{8} \left( 6 \cdot \mathrm{PSNR}_{\mathrm{Y}} + \mathrm{PSNR}_{\mathrm{U}} + \mathrm{PSNR}_{\mathrm{V}} \right).
\label{eq:psnr}
\end{equation}
The luma (Y) component is weighted by 6, which shall reflect the characteristic of the human perception that is more sensitive to luminance than to chrominance.

An alternative to evaluate the visual quality of video sequences is Video Multimethod Assessment Fusion (VMAF)~\cite{LiAaronKatsavounidisEtAl2016}, which is a full reference metric. According to~\cite{LiAaronKatsavounidisEtAl2016}, the subjective quality is predicted more accurately with VMAF than with PSNR. This is achieved by using a support vector machine that combines several quality metrics from the literature.

For the derivation of the VMAF score, the implementation of~\cite{Netflix2020} is used with the model version 0.6.1. The default VMAF score is trained on subjective viewing tests with full-HD displays. As we also consider sequences with a 4K resolution, we use the VMAF score that is trained for 4K displays for the corresponding video sequences.

In Figure~\ref{fig:BDDE}, several of the previously described metrics are visualized for the Tango2 input sequence. The diagram on the left side shows the $\mathrm{PSNR}_{\mathrm{YUV}}$ with the corresponding bit rate for the HM-16.20 encoder (blue) and the VTM-10.0 encoder (red). In the diagram in the middle, we show the decoding energy demand on the horizontal axis. Finally, the diagram on the right depicts the VMAF-score.

\subsection{Bj{\o}ntegaard-Delta Metrics}
\label{subsec:Bjontegaard}

In Section~\ref{sec:DSE}, we described both the metric BDR-PSNR and BDDE-PSNR. In the following, we also want to utilize VMAF, which has a higher correlation to the subjective impression of video sequences than PSNR. Therefore, we use the results of the VMAF score and substitute PSNR by VMAF for each previously mentioned BD metric to obtain BDR-VMAF and BDDE-VMAF. For BDR-VMAF, we obtain the bit rate savings for an equal VMAF score and for BDDE-VMAF, the decoder's energy savings, respectively. 

In Figure~\ref{fig:BDDE}, the diagram on the left shows that VVC has a higher compression efficiency because it uses less bit rate for equal $\mathrm{PSNR}_{\mathrm{YUV}}$ (BDR-PSNR compared to HM: \pro{-39.2}). For the diagram in the middle, the VTM decoder has a higher energy demand than the HM decoder, which results in a BDDE-PSNR of \pro{81.2} compared to HM decoding. Finally, for the diagram on the right side, the VTM decoder also has a higher energy demand than the the HM decoder with a BDDE-VMAF of \pro{79.9}.

\subsection{Efficiency Analysis of HEVC and VVC}
\label{subsec:HEVCvsVVC}

\begin{figure}[t!]
\begin{small}
\begin{center}
\begin{tikzpicture}
  \begin{axis}[
    width=0.47\textwidth,
    height = 0.4\textwidth,
      xlabel={BDR-VMAF in \%},
      ylabel={BDDE-VMAF in \%},
      axis lines = left,
      xmin=-40, xmax=2,
      ytick={0,20,40,60,80,100,120,140,160,180},
    ymin=0, ymax=192,    
      xtick={0,-5,-10,-15,-20,-25,-30,-35,-40},
      grid=both,
  xminorgrids=true,
  yminorgrids=true,
      ymajorgrids=true,
      xmajorgrids=true,
     grid style=dashed,
      legend cell align={left}, 
  ]
    \hspace*{-0.4cm}

  \addplot[only marks,
     color=black,
      mark=o,
      line width=1.5pt,
      mark size = 3pt,
      ]
      coordinates {
    (-38.52,86.17)
          };
      \addlegendentry{VTM JVET}
  
  \addplot[only marks,
     color=black,
      mark=diamond,
      line width=1.5pt,
      mark size = 3pt,
      ]
      coordinates {
    (-32.33,69.19)
      };
                  \addlegendentry{VTM UVG}

      \addplot[only marks,
     color=black,
      mark=+,
      line width=1.5pt,
      mark size = 3pt,
      ]
      coordinates {
    (-38.52,98.40) 
      };
      \addlegendentry{VVdeC JVET}

      \addplot[only marks,
     color=black,
      mark=asterisk,
      line width=1.5pt,
      mark size = 3pt,
      ]
      coordinates {
    (-32.33,51.00)
      };
              \addlegendentry{VVdeC UVG}
              
                  \addplot[only marks,
     color=black,
      mark=square*,
      line width=1.5pt,
          mark size = 2pt,
      ]
      coordinates {
      (0,0)
      };
              \addlegendentry{HEVC Decoder}    
  
  \addplot[only marks,
     color=blue,
      mark=o,
      line width=1.5pt,
      mark size = 3pt,
      ]
      coordinates {
      (-28.25,73.55)
            };
      
  \addplot[only marks,
     color=blue,
      mark=+,
      line width=1.5pt,
      mark size = 3pt,
      ]
      coordinates {
      (-28.25,174.57)
      };
      
      \addplot[only marks,
     color=red,
      mark=o,
      line width=1.5pt,
          mark size = 3pt,
      ]
      coordinates {
      (-31.92,75.65) 
      };

      \addplot[only marks,
     color=red,
      mark=+,
      line width=1.5pt,
          mark size = 3pt,
      ]
      coordinates {
      (-31.92,96.52)
      };

  \addplot[only marks,
     color=blue,
      mark=diamond,
      line width=1.5pt,
      mark size = 3pt,
      ]
      coordinates {
      (-26.22,75.51)
      };

  \addplot[only marks,
     color=blue,
      mark=asterisk,
      line width=1.5pt,
      mark size = 3pt,
      ]
      coordinates {
      (-26.22,171.32)
      };

  \addplot[only marks,
     color=red,
      mark=diamond,
      line width=1.5pt,
          mark size = 3pt,
      ]
      coordinates {
      (-26.46,61.48)
      };

      \addplot[only marks,
     color=red,
      mark=asterisk,
      line width=1.5pt,
          mark size = 3pt,
      ]
      coordinates {
      (-26.46,33.43) 
      };
      
    \node[black] at (axis cs: -36,98) {RA};
    \node[red]   at (axis cs: -29.5,96) {LB};
    \node[blue]  at (axis cs: -23.5,76) {AI};

  \end{axis}

\end{tikzpicture}
\end{center}
\end{small}
\vspace*{-0.4cm}

\caption{Evaluation of the compression and energy efficiency of both VVC decoders in relation to the corresponding decoders of HEVC. The horizontal axis shows the BDR-VMAF value and the vertical axis BDDE-VMAF. The blue markers correspond to the AI configuration, the red markers to LB, and the black markers to RA. The circle-shaped markers correspond to the average value of the JVET set with the VTM decoder, the diamond-shaped markers to the UVG set with VTM decoding, the plus-shaped marker to the JVET set with VVdeC decoding, and the asterisk-shaped markers to the UVG set with VVdeC decoding. The square-shaped marker corresponds to the reference of all coding configurations.}
\label{fig:HEVC_VVC}
\end{figure}
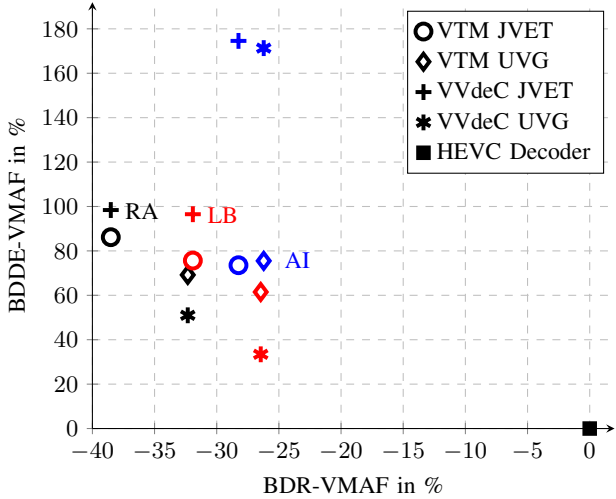

In the following, we will analyze the energy and compression efficiency of VVC compared to HEVC, which is shown in Figure~\ref{fig:HEVC_VVC}. In the figure, the horizontal axis depicts the average compression efficiency in terms of BDR-VMAF and the vertical axis the energy efficiency in terms of BDDE-VMAF for both test sets. For this figure, a low value for both metrics is desirable, which means that markers at the bottom left corner have a low increase in decoding energy demand in relation to the corresponding HEVC decoder and high decrease in bit rate. In the figure, we determine that the energy demand of the VTM decoder is increased in the range between \pro{60} and \pro{90} in relation to HM. For the optimized implementations, the energy demand increase of VVdeC is in the range between \pro{30} and \pro{190} in relation to openHEVC. In Table~\ref{tab:HEVCVVCComparison}, the BDR- and BDDE-values of the comparison of VVC in relation to HEVC are given. For the BDDE-values, we compare VTM with HM and VVdec with openHEVC. For each coding configuration, we show the results of each class and the mean average of each video sequence set, which will be discussed in the following.

\begin{table}[!t]
\def\arraystretch{1.05}
\caption{Evaluation of the compression and energy efficiency of HEVC and VVC. The compression efficiency is evaluated in terms of BDR-PSNR and BDR-VMAF and compares the VTM encoded bit streams in relation to HM. For the energy efficiency, we compare VTM and VVdeC in relation to HM and openHEVC, respectively. The additional energy demand of VVC is evaluated in terms of BDDE-PSNR and BDDE-VMAF.}
\label{tab:HEVCVVCComparison}
\begin{center}
\begin{tabular}{c || c : c | c : c | c : c } 
  & \multicolumn{2}{c|}{} 
  & \multicolumn{2}{c|}{VTM} 
  & \multicolumn{2}{c}{VVdeC}\\ 
  & \multicolumn{2}{c|}{BDR in $\%$} 
  & \multicolumn{2}{c|}{BDDE in $\%$} 
  & \multicolumn{2}{c}{BDDE in $\%$}\\ 
 &  PSNR & VMAF 
 &  PSNR & VMAF 
 &  PSNR & VMAF \\
 \hline \hline 
  & \multicolumn{6}{c}{All Intra} \\
 \hline 
	 A1 & -30.42 & -32.47 & 68.38 & 64.63 & 177.80 & 197.74 \\ 
     A2 & -28.57 & -31.76 & 79.65 & 74.05 & 151.99 & 169.41 \\ 
     B & -23.00 & -25.37 & 82.39 & 73.45 & 152.42 & 159.68 \\ 
     C & -22.28 & -23.35 & 92.97 & 84.67 & 170.18 & 178.64 \\ 
     D & -17.55 & -20.29 & 95.88 & 87.69 & 220.83 & 226.45 \\ 
     E & -25.68 & -26.35 & 67.88 & 64.56 & 194.10 & 203.01 \\ 
     F & -39.72 & -40.34 & 66.71 & 61.50 & 160.27 & 166.38 \\ 
       \hdashline
      JVET  & -26.43 & -28.25 & 80.07 & 73.55 & 174.57 & 184.42 \\ 
      \hline
     UVG1 & -22.83 & -25.56 & 69.96 & 65.98 & 168.35 & 185.82 \\ 
     UVG2 & -28.47 & -27.28 & 77.59 & 75.44 & 176.96 & 195.45 \\ 
     UVG3 & -26.34 & -25.75 & 81.96 & 76.63 & 168.27 & 178.30 \\ 
     \hdashline
      UVG  & -25.86 & -26.22 & 76.26 & 72.51 & 171.32 & 186.88 \\ 
       \hline \hline 
        & \multicolumn{6}{c}{Lowdelay B} \\
 \hline
     B & -31.81 & -32.06 & 74.86 & 73.53 & 27.51 & 27.44 \\ 
     C & -28.11 & -28.38 & 83.54 & 84.18 & 50.65 & 53.51 \\ 
     D & -24.44 & -25.74 & 91.88 & 93.42 & 203.46 & 245.19 \\ 
     E & -34.27 & -31.12 & 70.44 & 71.73 & 93.08 & 102.20 \\ 
     F & -43.12 & -42.04 & 54.44 & 54.96 & 124.32 & 128.10 \\ 
       \hdashline
       JVET  & -32.23 & -31.92 & 75.25 & 75.65 & 96.52 & 107.55 \\ 
      \hline
	UVG3  & -28.94 & -26.46 & 61.43 & 61.48 & 33.43 & 34.16 \\  
	 \hline \hline 
        & \multicolumn{6}{c}{Randomaccess} \\
 \hline 
	 A1 & -39.23 & -42.38 & 81.64 & 82.27 & 54.49 & 55.75 \\ 
     A2 & -42.87 & -49.26 & 100.62 & 96.82 & 45.12 & 42.83 \\ 
     B & -37.19 & -39.55 & 84.80 & 85.18 & 49.85 & 49.85 \\ 
     C & -30.36 & -30.69 & 85.51 & 86.65 & 103.29 & 109.96 \\ 
     D & -27.55 & -30.43 & 89.40 & 89.44 & 159.58 & 164.82 \\ 
     F & -42.46 & -42.22 & 78.16 & 78.59 & 165.91 & 171.56 \\ 
       \hdashline
      JVET  & -36.25 & -38.52 & 86.22 & 86.17 & 98.40 & 101.32 \\  
      \hline
     UVG1 & -32.37 & -28.99 & 63.78 & 68.33 & 59.72 & 56.97 \\ 
     UVG2 & -38.57 & -38.39 & 61.62 & 62.46 & 28.84 & 29.00 \\ 
     UVG3 & -31.23 & -29.21 & 76.99 & 77.87 & 66.35 & 68.30 \\ 
     \hdashline
      UVG  & -34.18 & -32.33 & 67.05 & 69.19 & 51.00 & 50.69 \\ 
 \hline \hline 
\end{tabular}
\end{center}
\vspace{-0.5cm}
\end{table}

For AI coding in Figure~\ref{fig:HEVC_VVC} (blue markers), we measured that the VTM decoder has an average energy demand increase of \pro{73.55} for the JVET set in terms of BDDE-VMAF, and by \pro{75.51} for the UVG set. For RA coding (black markers), the corresponding results are \pro{86.17} and \pro{69.19}, respectively. For LB coding (red markers), the average BDDE value is \pro{75.65} for the JVET set and \pro{61.48} for the UVG set. 

For VVdeC decoding, we observe that the increase of the energy demand is higher for AI coding compared to the VTM decoder. For the JVET set, BDDE-VMAF is \pro{174.57} and for the UVG set, \pro{171.32}. Furthermore, we determine that for the JVET set, the energy demand increase is also higher with the VVdeC decoder, which has a BDDE-VMAF of \pro{96.52} for LB coding, and of \pro{98.40} for RA coding. However, for the UVG set, the energy demand is lower with \pro{33.43} for LB coding and \pro{51.00} for RA coding.
By comparing the results of RA and LB for VVdeC for the corresponding classes in Table~\mbox{\ref{tab:HEVCVVCComparison}}, we observe that the sequences with UHD (Class A1, A2, UVG1, and UVG2) and HD resolution (B and UVG3) have a lower BDDE value than the other classes (C, D, E, and F). Therefore, we conclude that video sequences with a higher resolution have a lower relative increase in energy demand than sequences with a lower resolution.

In summary, both VVC decoders have a significant increase in energy demand. Consequently, it is desirable to find coding tool profiles that have a lower energy demand than the reference profiles of VVC for all coding configurations.

\section{Evaluation}
\label{sec:evaluation}

In the following, we will at first evaluate the proposed algorithm on a subset of coding tools in Subsection~\ref{subsec:DSE_Eval}. Then, the results of the DSE algorithm and our tool sensitivity analysis will be presented in Subsection~\ref{subsec:Training}. Finally, in Subsection~\ref{subsec:Validation}, we will select two coding tool profiles~$\boldsymbol{u}_{\boldsymbol{i,\nu}}$ for AI, LB, and RA coding and validate them on both test sets with VTM and VVdeC. 

\subsection{DSE for Subset of Coding Tools}
\label{subsec:DSE_Eval}
As shown in Section~\ref{sec:DSE}, a full search is not possible for all coding tools. However, to evaluate the algorithm on optimality, we select a subset of coding tools for each coding configuration. For each subset, we randomly selected the following coding tools from the groups in Table~\ref{tab:ToolList}: ISP, CCLM, DQ, MTS, ALF, SAO, AFFINE, and GPM. 

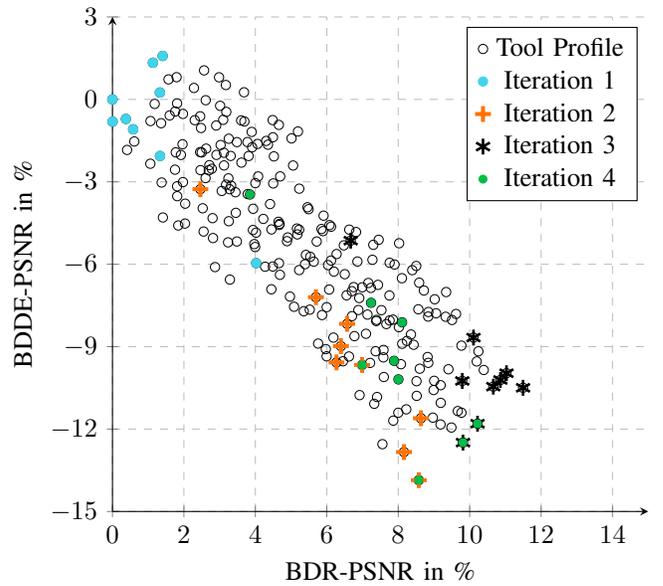
\begin{figure}[!t]
\definecolor{Iter1}{HTML}{40D4EB}
\definecolor{Iter2}{HTML}{FF7100}
\definecolor{Iter3}{HTML}{FFF100}
\definecolor{EE}{HTML}{00BD47}

\begin{center}
\begin{tikzpicture}
\begin{axis}[
	width=0.48\textwidth,
	height = 0.45\textwidth,
    xlabel={BDR-PSNR in \%},
    ylabel={BDDE-PSNR in \%},
    xmin=0, xmax=15,
    ymin=-15, ymax=3,
    ytick={3,0,-3,-6,-9,-12,-15},
        axis lines = left,
    ymajorgrids=true,
    xmajorgrids=true,
    grid style=dashed
    ]

\addplot[only marks,
   color=black,
    mark=o,
    mark size=1.75pt,
    line width=0.25pt,
    ]
    coordinates {
        (0,0)
        (0.37533,-0.71072)
        (1.3274,0.25003)
        (1.7558,-0.4402)
        (1.1324,1.335)
        (1.5692,0.72906)
        (2.5645,1.0563)
        (2.984,0.80447)
        (0.57906,-1.0892)
        (1.065,-0.78885)
        (1.9731,-0.52845)
        (2.4547,-0.6163)
        (1.795,0.80794)
        (2.2807,0.41621)
        (3.1744,0.52331)
        (3.6828,0.2586)
        (4.0233,-5.9585)
        (4.4893,-6.083)
        (5.5376,-5.7293)
        (6.0357,-6.0305)
        (5.2274,-4.7437)
        (5.7172,-5.1738)
        (6.8193,-4.8286)
        (7.3019,-5.2937)
        (4.6614,-5.9615)
        (5.2268,-6.2128)
        (6.1992,-6.2679)
        (6.7524,-6.833)
        (5.8909,-4.7948)
        (6.4773,-5.3131)
        (7.4533,-5.1102)
        (8.025,-5.2371)
        (0.0034756,-0.81062)
        (0.4092,-1.8397)
        (1.362,-0.86735)
        (1.7725,-1.9273)
        (1.1814,-0.16434)
        (1.6012,-0.58383)
        (2.6103,-0.029942)
        (3.0138,-0.23364)
        (0.61167,-1.532)
        (1.0604,-2.3416)
        (1.9615,-1.633)
        (2.454,-1.9186)
        (1.8085,-0.14555)
        (2.3038,-0.7568)
        (3.1869,-0.45421)
        (3.6938,-0.80419)
        (4.4485,-6.9128)
        (4.969,-7.1781)
        (6.0667,-6.9221)
        (6.5549,-7.1309)
        (5.6351,-5.9059)
        (6.1189,-5.8664)
        (7.2779,-5.8393)
        (7.7599,-5.9476)
        (5.0837,-7.5481)
        (5.6959,-7.2)
        (6.7014,-6.9815)
        (7.2401,-7.4035)
        (6.2749,-5.6297)
        (6.8579,-5.9013)
        (7.8975,-6.0111)
        (8.5032,-6.064)
        (1.3318,-2.0598)
        (1.7515,-2.5582)
        (2.6953,-2.5801)
        (3.123,-2.4035)
        (2.4794,-1.0414)
        (2.9038,-0.94926)
        (3.8973,-1.1836)
        (4.3043,-1.5106)
        (1.9157,-2.0227)
        (2.4639,-2.5746)
        (3.3094,-2.0583)
        (3.8119,-2.0608)
        (3.0998,-0.74264)
        (3.6034,-0.76594)
        (4.4849,-0.75334)
        (5.0126,-1.4198)
        (5.3553,-7.7109)
        (5.8675,-7.8498)
        (6.8834,-7.667)
        (7.3837,-8.0689)
        (6.5033,-6.3634)
        (6.9928,-6.8394)
        (8.1031,-6.5088)
        (8.5858,-6.6719)
        (5.9962,-7.5631)
        (6.5704,-8.1711)
        (7.5291,-7.6724)
        (8.1073,-8.1155)
        (7.1622,-6.6822)
        (7.7542,-7.1385)
        (8.718,-6.9564)
        (9.3407,-7.013)
        (1.3771,-3.0254)
        (1.8128,-3.5231)
        (2.7124,-2.3931)
        (3.1804,-3.1035)
        (2.4979,-1.9312)
        (2.9108,-2.0203)
        (3.9093,-1.7431)
        (4.3535,-1.6494)
        (1.9736,-3.0109)
        (2.4612,-3.2679)
        (3.323,-3.1405)
        (3.8492,-3.4615)
        (3.1194,-1.9421)
        (3.6105,-2.3012)
        (4.5189,-2.0749)
        (5.057,-2.4067)
        (5.7573,-8.8883)
        (6.2717,-9.5747)
        (7.3676,-9.1673)
        (7.8823,-9.5152)
        (6.9091,-7.2315)
        (7.3931,-8.0264)
        (8.5633,-7.641)
        (9.0597,-7.7855)
        (6.3955,-8.9812)
        (6.9892,-9.6677)
        (8.0054,-10.1922)
        (8.5756,-13.8607)
        (7.5575,-12.55)
        (8.1604,-12.837)
        (9.1848,-11.8327)
        (9.8124,-12.4917)
        (1.4056,1.5832)
        (1.8093,-3.1635)
        (2.7796,-2.5245)
        (3.2457,-2.8736)
        (2.6223,-1.7926)
        (3.0276,-1.6609)
        (4.0609,-1.621)
        (4.4925,-1.3783)
        (2.0364,-3.7958)
        (2.526,-3.9636)
        (3.4435,-3.3797)
        (3.9382,-3.9815)
        (3.2558,-1.4321)
        (3.778,-1.5528)
        (4.6578,-0.92801)
        (5.1933,-1.1675)
        (5.5223,-7.654)
        (5.9952,-9.3524)
        (7.0972,-8.5319)
        (7.5839,-9.3913)
        (6.7056,-7.8907)
        (7.2349,-6.862)
        (8.3926,-6.2154)
        (8.844,-7.7387)
        (6.186,-8.6701)
        (6.774,-8.6185)
        (7.7588,-8.0471)
        (8.3556,-8.557)
        (7.433,-6.7516)
        (8.0189,-8.2917)
        (9.0209,-7.2758)
        (9.6174,-7.8089)
        (1.4116,-4.2984)
        (1.8409,-4.5892)
        (2.8025,-4.3228)
        (3.2572,-4.4347)
        (2.6067,-2.8641)
        (3.0591,-3.333)
        (4.0785,-2.8053)
        (4.5003,-3.0591)
        (2.031,-4.5202)
        (2.5512,-4.8099)
        (3.4532,-4.1288)
        (3.9324,-4.7549)
        (3.2816,-3.2696)
        (3.7861,-3.4458)
        (4.6954,-2.803)
        (5.2293,-3.2501)
        (5.9723,-9.0901)
        (6.4525,-9.5329)
        (7.5885,-9.1209)
        (8.0754,-9.6394)
        (7.1391,-7.8334)
        (7.668,-8.1778)
        (8.8379,-7.9058)
        (9.3276,-7.9121)
        (6.6207,-9.358)
        (7.2119,-9.5928)
        (8.2425,-8.7925)
        (8.8221,-9.3769)
        (7.8529,-8.0156)
        (8.4315,-8.9313)
        (9.5057,-8.0294)
        (10.1071,-8.6701)
        (2.815,-5.0483)
        (3.2361,-5.2037)
        (4.2229,-4.3502)
        (4.6793,-4.5308)
        (3.9996,-3.0487)
        (4.4113,-3.8108)
        (5.4601,-3.9261)
        (5.8521,-4.7182)
        (3.4836,-1.5551)
        (3.97,-5.2676)
        (4.8863,-5.1618)
        (5.3912,-5.9985)
        (4.6473,-4.4547)
        (5.1639,-4.7001)
        (6.0765,-4.5757)
        (6.6292,-4.7412)
        (6.9184,-10.7647)
        (7.4073,-10.833)
        (8.5166,-10.7917)
        (8.9907,-10.5801)
        (8.111,-8.8963)
        (8.6026,-9.8033)
        (9.7693,-8.9632)
        (10.2445,-9.1694)
        (7.597,-10.1054)
        (8.2242,-11.2826)
        (9.19,-11.0466)
        (9.7693,-11.4049)
        (8.8412,-9.589)
        (9.4057,-9.823)
        (10.3941,-9.8496)
        (11.0305,-9.9671)
        (2.8708,-6.1035)
        (3.2874,-6.5599)
        (4.2728,-6.0899)
        (4.6882,-6.3853)
        (4.019,-4.8516)
        (4.4514,-4.734)
        (5.4914,-3.9969)
        (5.9102,-4.2322)
        (3.482,-5.1718)
        (3.9947,-5.3736)
        (4.9074,-5.0248)
        (5.4157,-5.6597)
        (4.6671,-4.5713)
        (5.1762,-4.9143)
        (6.1201,-4.6339)
        (6.6698,-5.1425)
        (7.3286,-11.08)
        (7.8614,-11.6933)
        (8.9686,-11.5845)
        (9.489,-11.944)
        (8.5121,-10.2569)
        (9.0059,-10.2434)
        (10.1968,-9.5382)
        (10.6612,-10.4557)
        (8.0016,-11.4057)
        (8.6359,-11.6063)
        (9.6675,-11.3455)
        (10.2197,-11.8039)
        (9.1873,-10.4033)
        (9.7925,-10.2488)
        (10.8602,-10.2132)
        (11.4945,-10.4974)
    };
    \addlegendentry{Tool Profile}

\addplot[only marks,
   color=Iter1,
       fill=Iter1,
    mark=*,
    mark size=1.75pt,
    line width=0.1pt,
    ]
    coordinates {  
    (0,0)
    (0.37533,-0.71072)
    (1.3274,0.25003)
    (1.1324,1.335)
    (0.57906,-1.0892)
    (4.0233,-5.9585)
    (0.0034756,-0.81062)
    (1.3318,-2.0598)
    (1.4056,1.5832)
    };
    \addlegendentry{Iteration 1}

\addplot[only marks,
   color=Iter2,
   fill=Iter2,
    mark=+,
    mark size=3pt,
    line width=1.5pt,
    ]
    coordinates {
        (2.4612,-3.2679)
        (5.6959,-7.2)
        (6.5704,-8.1711)
        (6.3955,-8.9812)
        (6.2717,-9.5747)
        (6.9892,-9.6677)
        (8.6359,-11.6063)
        (8.1604,-12.837)
        (8.5756,-13.8607)
    };
    \addlegendentry{Iteration 2}

\addplot[only marks,
    color=black,
    fill=black,
    mark=asterisk,
    mark size=3pt,
    line width=1pt,
    ]
    coordinates {
        (6.6698,-5.1425)
        (10.1071,-8.6701)
        (11.0305,-9.9671)
        (10.8602,-10.2132)
        (9.7925,-10.2488)
        (10.6612,-10.4557)
        (11.4945,-10.4974)
        (10.2197,-11.8039)
        (9.8124,-12.4917)
    };
    \addlegendentry{Iteration 3}

\addplot[only marks,
    color=EE,
    fill=EE,
    mark=*,
    mark size=1.5pt,
    line width=0.1pt,
    ]
    coordinates {
    (3.8492,-3.4615)  
    (7.2401,-7.4035)  
    (8.1073,-8.1155)  
    (7.8823,-9.5152)  
    (6.9892,-9.6677)   
    (8.0054,-10.1922) 
    (10.2197,-11.8039) 
    (9.8124,-12.4917)  
    (8.5756,-13.8607)  
    };
    \addlegendentry{Iteration 4}    
    
\end{axis}
	
\end{tikzpicture}
\end{center}

\caption{ Results of all 256 coding tool profiles for a subset of 8 coding tools for RA. The additional bit rate (BDR-PSNR) is shown on the horizontal axis and the energy saving are shown by the vertical axis (BDDE-PSNR). Each coding tool profile corresponds to a black circlce. Each iteration of the greedy algorithm is shown by markers in different colors.}
\label{fig:Greedy}
\end{figure}

In the following, we encode all combinations of the corresponding coding tools for the class C sequences. For RA and LB, we encode 256 coding tool profiles and for AI, 64 profiles since the coding tools AFFINE and GPM are not applicable. Figure~\ref{fig:Greedy} shows the results of the evaluation with the greedy algorithm for RA coding. In the figure, the vertical axis shows the BDDE-PSNR and the horizontal axis the BDR-PSNR. For the visualization of Algorithm~\ref{alg:Gradient} in Figure~\ref{fig:Greedy}, we use different colors for each iteration. We determine from Figure~\ref{fig:Greedy} that the algorithm successfully obtained the coding tool profile with the highest energy demand reduction in terms of BDDE-PSNR, which is \pro{-13.86}. This coding tool profile is determined both in the second iteration (orange marker) and the forth iteration (green marker).

For LB and AI, the algorithm also successfully determined the coding tool profile with the lowest BDDE-PSNR value. Thus, although we are not able to prove optimality for all coding tools, we show that the optimal solution was determined for each subset of coding tools.

 \subsection{Training \& Tool Sensitivity Analysis}
 \label{subsec:Training}

\begin{figure}[!t]
\definecolor{Iter1}{HTML}{40D4EB}
\definecolor{Iter2}{HTML}{FF7100}
\definecolor{Iter3}{HTML}{FFF100}
\definecolor{EE}{HTML}{00BD47}

\begin{center}
\begin{tikzpicture}
\begin{groupplot}[
     group style={group name=my plots,group size= 1 by 3,vertical sep =2.25cm},
          title style={yshift=-0.25cm},
     height = 10cm,
     width = \textwidth,
    ]
    
    \nextgroupplot[
	width=0.48\textwidth,
	height = 0.36\textwidth,
    xlabel={BDR-PSNR in $\%$ },
    title = {(a) All Intra},
    ylabel={BDDE-PSNR in $\%$},
    xmin=-1, xmax=32,
    ymin=-50, ymax=12,
    ytick={10,0,-10,-20,-30,-40,-50},
        axis lines = left,
    ymajorgrids=true,
    xmajorgrids=true,
    grid style=dashed,
]

\addplot[only marks,
   color=black,
   fill=Iter1,
    mark=*,
    mark size=1.75pt,
    line width=0.25pt,
    ]
    coordinates {
    (2.3015,   -3.4978)
   (-0.0187,    0.2025)
   ( 0.2747,   -0.9890)
   ( 2.3925,    0.0441)
   ( 1.1053,    5.6716)
   (-0.3989,  -17.2878)
   ( 1.2224,    3.0770)
   (-0.3912,    1.5061)
   ( 0.7022,   -1.6529)
   ( 0.5483,    1.3694)
   ( 1.2010,   -1.0984)
   (-0.1010,   -0.9498)
   ( 0.4736,    0.0584)
   ( 0.5999,    1.2321)
   ( 0.7452,    0.1001)
   ( 0.0471,   -2.0060)
    };
    \addlegendentry{Iteration 1}

\addplot[only marks,
   color=black,
       fill=Iter2,
    mark=*,
    mark size=1.75pt,
    line width=0.25pt,
    ]
    coordinates {
    (5.1355,  -26.5203)
    (1.6358,  -23.9748)
    (5.0847,  -26.4224)
    (5.1355,  -26.7855)
    (7.9219,  -28.0772)
    (6.5424,  -23.5913)
    (4.6404,   -3.9057)
    (6.3187,  -25.1620)
    (4.7193,  -25.9582)
    (4.5313,  -25.5182)
    (5.7858,  -26.3690)
    (3.8031,  -26.1150)
    (5.1084,  -25.1379)
    (5.8956,  -27.2369)
    (6.1054,  -26.4172)
    (7.4537,  -28.6913)
    (4.3985,  -24.6420)
    };
    \addlegendentry{Iteration 2}

\addplot[only marks,
   color=black,
   fill=Iter3,
    mark=*,
    mark size=1.75pt,
    line width=0.25pt,
    ]
    coordinates {
   (11.2397,  -30.9300)
   ( 6.4233,  -27.1038)
   (11.0979,  -30.6243)
   (11.2397,  -29.9854)
   ( 8.4911,  -29.5076)
   (12.3975,  -28.4240)
   (10.0462,   -7.7013)
   (13.0093,  -28.7801)
   (10.8332,  -30.2827)
   (10.4126,  -29.3814)
   (12.4518,  -30.3054)
   ( 8.1241,  -28.2712)
   (11.0795,  -29.1456)
   (10.2679,  -30.3550)
   (12.4184,  -30.2563)
   ( 8.6190,  -29.1728)
   (10.1956,  -29.0321)
    };
    \addlegendentry{Iteration 3}

\addplot[only marks,
   color=blue,
    mark=+,
        mark size=3pt,
    line width=1.5pt,
    ]
    coordinates {
    (29.72,-48.11)
    };
    \addlegendentry{HM CTC}

\addplot[only marks,
   color=red,
    mark=+,
    line width=1.5pt,
        mark size = 3pt,
    ]
    coordinates {
    (0,0)
    };
    \addlegendentry{VTM CTC}    
    
    \addplot[only marks,
   draw=black,
   color=EE,
    mark=asterisk,
    line width=1.5pt,
    mark size =3pt,
    ]
    coordinates {
    (3.8031,  -26.1150)  
    };
    \addlegendentry{VTM EBE}

\addplot[only marks,
   draw=black,
   color=EE,
    mark=diamond*,
    mark size =3pt,
    line width=1.5pt,
    ]
    coordinates {
   (11.2397,  -30.9300)
    };
    \addlegendentry{VTM EE}

\nextgroupplot[
	width=0.48\textwidth,
	height = 0.36\textwidth,
    xlabel={BDR-PSNR in $\%$},
    title = {(b) Lowdelay B},
    ylabel={BDDE-PSNR in $\%$},
    xmin=-1, xmax=45,
    ymin=-50, ymax=12,
        ytick={10,0,-10,-20,-30,-40,-50},
        axis lines = left,
    ymajorgrids=true,
    xmajorgrids=true,
    grid style=dashed,
    yshift=0.33cm,
]

\addplot[only marks,
   color=black,
   fill=Iter1,
    mark=*,
    mark size=1.75pt,
    line width=0.25pt,
    ]
    coordinates {
    (     0,         0)
    (1.5385,   -5.1052)
    (3.9879,   -6.1280)
    (0.5899,    0.8598)
    (0.0900,    0.0789)
    (-0.0669,    0.4527)
    (0.2386,   -0.8543)
    (0.4865,    0.0275)
    (0.3386,   -0.6933)
    (0.1307,    0.3161)
    (0.8621,  -20.0386)
    (1.3585,    1.5365)
    (2.2401,    4.2694)
   (-0.0902,    0.0197)
   ( 0.0518,    0.2083)
   ( 0.2446,    0.4681)
   (-0.5391,    0.5340)
   ( 0.2186,   -2.1417)
   (-0.2752,   -0.0578)
   ( 0.4076,    1.1231)
   ( 0.1007,    0.1107)
   ( 0.4342,   -0.8992)
   ( 0.1650,   -0.3136)
   ( 0.0859,   -1.2310)
   ( 0.4641,   -0.8964)
   ( 0.2625,   -1.9487)
    };
    \addlegendentry{Iteration 1}

\addplot[only marks,
   color=black,
   fill=Iter2,
    mark=*,
    mark size=1.75pt,
    line width=0.25pt,
    ]
    coordinates {
   (10.7515,  -38.4471)
   ( 9.6963,  -34.8092)
   ( 4.6860,  -32.6991)
   (12.1923,  -37.8536)
   (11.3573,  -38.8923)
   (10.6857,  -38.4978)
   (10.7515,  -38.6272)
   (11.5396,  -38.6600)
   (10.2108,  -37.4931)
   (10.9738,  -38.6293)
   ( 7.9641,  -20.3850)
   (12.2094,  -37.2408)
   (13.7219,  -37.3441)
   (10.7980,  -39.3747)
   (10.9844,  -39.8843)
   (11.1010,  -39.5238)
   ( 9.9147,  -38.6802)
   (10.2961,  -36.2714)
   (11.0825,  -39.2638)
   (11.5947,  -38.6378)
   (11.1391,  -38.8184)
   (10.1434,  -38.7318)
   (10.7515,  -39.2115)
   ( 8.9955,  -37.0099)
   (10.3735,  -38.8694)
   (10.5849,  -37.4846)
    };
    \addlegendentry{Iteration 2}

\addplot[only marks,
   color=black,
   fill=Iter3,
    mark=*,
    mark size=1.75pt,
    line width=0.25pt,
    ]
    coordinates {
   (12.5844,  -38.1322)
   (11.0571,  -33.1693)
   ( 5.5317,  -31.5811)
   (14.4363,  -37.7728)
   (12.0492,  -37.7788)
   (12.5409,  -38.2526)
   (12.5844,  -38.0861)
   (11.6727,  -38.2179)
   (12.1323,  -37.3955)
   (12.2745,  -39.4834)
   ( 9.9844,  -20.1494)
   (13.9179,  -36.8614)
   (15.6554,  -36.2495)
   (12.5154,  -38.0864)
   (12.1829,  -38.2352)
   (12.1664,  -38.0823)
   (13.1311,  -38.4819)
   (11.9942,  -35.4144)
   (12.3213,  -37.9242)
   (11.6530,  -38.7700)
   (12.3007,  -37.8892)
   (13.3647,  -38.6420)
   (12.5844,  -37.8690)
   (10.9249,  -35.6563)
   (13.3883,  -39.1997)
   (12.4167,  -36.6148)
    };
    \addlegendentry{Iteration 3}

\addplot[only marks,
   color=blue,
    mark=+,
    line width=1.5pt,
        mark size = 3pt,
    ]
    coordinates {
    (40.27,-45.15)
    };
    \addlegendentry{HM CTC}

\addplot[only marks,
   color=red,
    mark=+,
        mark size=3pt,
    line width=1.5pt,
    ]
    coordinates {
    (0,0)
    };
    \addlegendentry{VTM CTC}
    
    \addplot[only marks,
   color=EE,
    mark=asterisk,
    mark size =3pt,
    line width=1.5pt,
    ]
    coordinates {
   ( 4.6860,  -32.6991)
    };
    \addlegendentry{VTM EBE}

\addplot[only marks,
   color=EE,
   fill=EE,
    mark=diamond*,
    mark size =3pt,
    line width=1.5pt,
    ]
    coordinates {
     (10.9844,  -39.8843)
    };
    \addlegendentry{VTM EE}
    
\nextgroupplot[
	width=0.48\textwidth,
	height = 0.36\textwidth,
    xlabel={BDR-PSNR in $\%$},
    ylabel={BDDE-PSNR in $\%$},
    title = {(c) Randomaccess},
    xmin=-1, xmax=50,
    ymin=-50, ymax=12,
        ytick={10,0,-10,-20,-30,-40,-50},
        axis lines = left,
    ymajorgrids=true,
    xmajorgrids=true,
    grid style=dashed,
    legend style={draw=none},
        yshift=0.33cm,
]

\addplot[only marks,
   color=black,
   fill=Iter1,
    mark=*,
    mark size=1.75pt,
    line width=0.25pt,
    ]
    coordinates {
   (      0 ,        0)
   ( 1.0173 ,  -1.8487)
   ( 3.6250 ,  -5.5685)
   ( 1.1467 ,   0.3906)
   ( 0.1447 ,   1.0640)
   ( 0.7125 ,  -2.5810)
   (-0.0415 ,  -0.1456)
   ( 0.1729 ,  -0.8735)
   ( 1.0680 ,  -0.1471)
   ( 0.2611 ,  -0.3036)
   ( 0.3278 ,   0.3394)
   ( 0.8907 , -12.8395)
   ( 0.4689 ,  -4.6101)
   ( 1.2940 ,   1.3292)
   ( 1.8402 ,   2.9079)
   ( 0.1425 ,  -0.3392)
   ( 0.3805 ,  -0.4279)
   ( 0.4746 ,   0.0946)
   ( 0.7577 ,  -0.7503)
   ( 0.6696 ,  -0.2526)
   ( 0.2204 ,  -0.4360)
   ( 0.3665 ,   0.5633)
   ( 0.2518 ,   0.1507)
   ( 0.5487 ,  -0.3326)
   ( 0.3630 ,  -0.3953)
   ( 0.0544 ,  -1.0790)
   ( 0.4115 ,  -0.8924)
   ( 0.4225 ,  -0.8860)
   ( 0.1932 ,   0.2582)
    };
    \addlegendentry{Iteration 1}

\addplot[only marks,
   color=black,
   fill=Iter2,
    mark=*,
    mark size=1.75pt,
    line width=0.25pt,
    ]
    coordinates {
   ( 16.6724,  -43.3986)
   ( 15.6242,  -40.1286)
   ( 9.9992 , -36.9296)
   ( 18.3938,  -43.2918)
   ( 17.0844,  -44.1319)
   ( 15.1281,  -34.6562)
   ( 16.7795,  -43.2631)
   ( 16.6724,  -43.5409)
   ( 15.3917,  -43.5097)
   ( 16.1853,  -43.1271)
   ( 17.0023,  -43.4998)
   ( 13.8816,  -27.2753)
   ( 15.7520,  -33.9519)
   ( 18.3128,  -43.0814)
   ( 19.7904,  -42.3357)
   (16.4272 , -43.6494)
   (16.2633 , -43.1936)
   (17.4135 , -43.8402)
   (15.0432 , -43.2042)
   (15.8729 , -43.2216)
   (16.1363 , -43.7620)
   (17.8069 , -43.5977)
   (17.2236 , -44.0104)
   (15.3567 , -43.6292)
   (16.2684 , -43.9905)
   (15.5019 , -42.3311)
   (16.2910 , -43.6489)
   (16.1202 , -41.4246)
   (16.6724 , -43.7533)
    };
    \addlegendentry{Iteration 2}

\addplot[only marks,
   color=black,
   fill=Iter3,
    mark=*,
    mark size=1.75pt,
    line width=0.25pt,
    ]
    coordinates {
   (14.6113 , -39.6155)
   (12.9596 , -35.4245)
   ( 8.2143 , -32.5607)
   (16.2168 , -39.2724)
   (14.1570 , -39.1390)
   (13.1357 , -29.7085)
   (14.6375 , -39.7384)
   (14.6113 , -39.6251)
   (16.2976 , -40.0070)
   (14.3004 , -39.1529)
   (14.2465 , -39.8959)
   (12.4730 , -22.0383)
   (13.5601 , -28.3395)
   (15.9492 , -39.2165)
   (17.5821 , -38.4661)
   (14.7743 , -39.9644)
   (14.1233 , -39.5199)
   (14.0858 , -39.9246)
   (13.7366 , -39.5974)
   (13.9369 , -39.6060)
   (15.1931 , -40.1238)
   (13.8964 , -40.2976)
   (14.1516 , -40.3179)
   (16.3257 , -40.8600)
   (15.3733 , -39.9282)
   (13.5919 , -38.2137)
   (15.2212 , -40.8681)
   (13.9963 , -37.5050)
   (15.3733 , -40.0484)
    };
    \addlegendentry{Iteration 3}

\addplot[only marks,
   color=blue,
    mark=+,
        mark size=3pt,
    line width=1.5pt,
    ]
    coordinates {
    (43.89,-46.02)
    };
    \addlegendentry{HM CTC}

\addplot[only marks,
   color=red,
    mark=+,
        mark size=3pt,
    line width=1.5pt,
    ]
    coordinates {
    (0,0)
    };
    \addlegendentry{VTM CTC}
    
\addplot[only marks,
   color=EE,
    mark=asterisk,    
    mark size =3pt,
    line width=1.5pt,
    ]
    coordinates {
  ( 9.9992 , -36.9296)    
    };
    \addlegendentry{VTM EBE}

\addplot[only marks,
   color=EE,
   fill=EE,
    mark=diamond*,    
    mark size =3pt,
    line width=1.5pt,
    ]
    coordinates {
   ( 17.0844,  -44.1319) 
    };
    \addlegendentry{VTM EE}

\end{groupplot}

\end{tikzpicture}
\end{center}

\caption{Training results of the DSE algorithm for each CTC configuration. In (a), the results of AI coding are shown, in (b) for LB, and in (c) for RA. The horizontal axis shows the additional bit rate measured in BDR-PSNR and the vertical axis shows the energy demand reduction measured in BDDE-PSNR. Additionally, for each CTC configuration, two reference points are given, which are VTM decoding with CTC profile encoded bit streams (red plus-shaped marker) and HM decoding with CTC profile encoded bit streams (blue plus-shaped marker). Furthermore, the results of each iteration are depicted with circular markers.}
\label{fig:Training}
\end{figure}
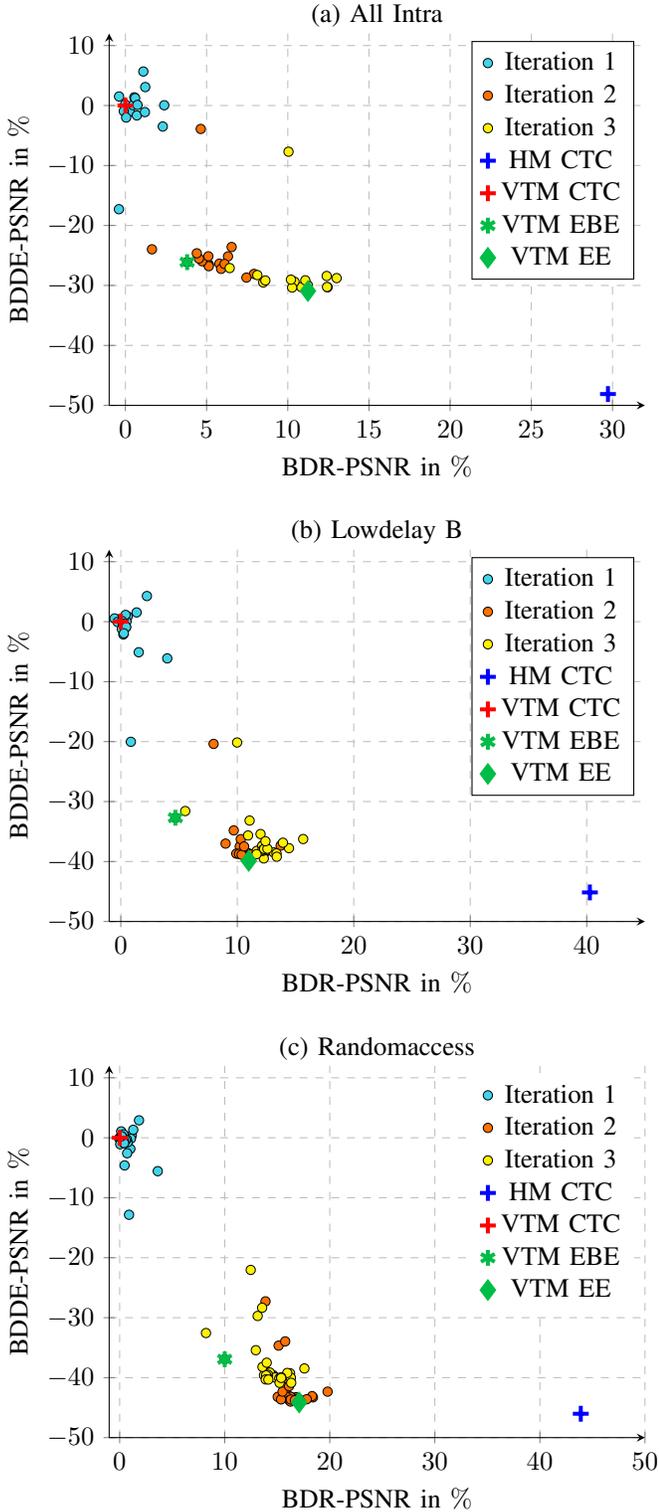

For the training of the DSE, we only use class C sequences from the JVET set, which has sufficient coverage of the Spatial Information (SI) and Temporal Information (TI) according to \cite{Amestoy2020}. Furthermore, we reduce the number of encoded frames to 128. Thereby, we reduce the time demand and computational complexity to encode bit streams. The coding tools in Table~\ref{tab:ToolList} are selected based on the recommendations in~\cite{JVET-T0013}. If the coding tool dependent quantization (DQ) is disabled, we enable the coding tool sign data hiding, which is an alternative to DQ. Furthermore, we use VTM-8.0 as an encoder, which has the same coding tool set as VTM-10.0.

In Figure~\ref{fig:Training}, the results for the training of all coding tool profiles are shown. The red markers correspond to the performance of the CTC profile of VTM, which will be the reference for the calculation of the BDR and BDDE metrics in the following. The blue plus-shaped markers correspond to the results of the HEVC CTC profile. Additionally, for each CTC coding configuration, two profiles are selected that will be evaluated profoundly in Section~\ref{subsec:Validation}. The green diamond-shaped marker corresponds to the profile with maximum energy reduction, which we call energy efficient (EE) profile. The green asterisk marker corresponds to the joint tradeoff of the energy demand reduction and the bit rate increase, which we call energy and bit rate efficient (EBE). We select the EBE profile for each coding configuration based on a refinement selection. At first, we select the best three coding tool profiles with a BDR value of less than \pro{10} and the lowest sum of BDDE and BDR. Then, we encode and measure five sequences of the JVET set. Finally, the profile with the lowest sum of BDDE and BDR after the refinement selection will be used as the EBE profile.

Additionally, we will analyze the energy sensitivity of all tested coding tools. Therefore, according to their influence on the energy efficiency for the first iteration of the DSE, we assign each coding tool to one of the following four categories: major increase, minor increase, minor decrease, major decrease. Coding tools that change the energy demand by their usage by $\pm$\pro{1} will be assigned to the categories with minor influence. Since most of the coding tools in Table~\ref{tab:ToolList} are initially used in the CTC profile, the DSE disables each coding tool separately in the first iteration. Therefore, a positive BDDE value corresponds to higher energy demand when the corresponding coding tool is not used. Consequently, the use of the coding tool improves the energy efficiency. The remaining coding tools, which are initially disabled in Table~\ref{tab:ToolList}, are enabled in the first iteration of the DSE, and a negative BDDE value corresponds to a lower energy demand when the corresponding coding tool is used. The results are summarized in Table~\ref{tab:CodingToolDescription}.

\subsubsection{All Intra}

For AI coding (cf. Fig.~\ref{fig:Training} (a)), the deblocking filter has the highest energy demand reduction within the first iteration with a BDDE-PSNR value of \pro{-17.29} and a BDR-PSNR of \pro{-0.40}, which corresponds to the blue marker. Therefore, it can be concluded that DBF has a huge potential to optimize the energy demand of the decoder, which is similar to the findings of~\cite{NoguesMenardPelcat2019}, where the deblocking filter was skipped.

For the second iteration, we can see in Figure~\ref{fig:Training}~(a) that the updated coding tool profile~$\boldsymbol{u}_{\boldsymbol{2,0}}$ has a BDDE-PSNR value of \pro{-26.52} and a BDR-PSNR value of \pro{5.14}. For this profile, all coding tools that have a lower BDDE-PSNR value than \pro{0} are disabled, which is fulfilled for the tools ALF, CCALF, DBF, ISP, LFNST, LMCS, and SAO.

Finally, in the third iteration, the lowest BDDE-PSNR value is measured for the coding tool profile~$\boldsymbol{u}_{\boldsymbol{3,0}}$ with \pro{-30.93} and a BDR-PSNR value of \pro{11.24}. This coding tool profile will be used as the EE profile, which is also visualized by the diamond-shaped marker in Figure~\ref{fig:Training}. Furthermore, the usage of each coding tool is shown in Table~\ref{tab:EnergyEfficientConfigs}. For the other profiles in the third iteration, we determined that a change in the usage decreases the energy efficiency for all coding tools. Therefore, the condition for the termination of Algorithm~\ref{alg:Gradient} is met. 

The green asterisk marker in Figure~\ref{fig:Training}~(a) corresponds to the EBE profile, which is a profile that we selected from iteration 2 based on the refinement that we explained above. Again, the usage of the corresponding coding tools is shown in Table~\ref{tab:EnergyEfficientConfigs}. The BDDE-PSNR value for this profile is \pro{-25.96} and the BDR-PSNR value is \pro{4.72}.

\begin{table}[t!]
\def\arraystretch{1.05}
\caption{Results of the coding tool sensitivity analysis. All coding tools are assigned to one of four categories for each coding configuration according to their influence on the energy efficiency of the VTM decoder if a coding tool is disabled.}
\vspace*{-0.4cm}
\label{tab:CodingToolDescription}
\begin{center}
\begin{tabular}{ l | l   } 
 & AI  \\
\hline\hline
Major Increase & CST, DQ, JCCR, MRL  \\
\hdashline
Minor Increase & CCLM, MIP, MTS   \\
\hline 
Major Decrease & ALF, DBF, IBC, ISP, LFNST, SAO  \\
\hdashline
Minor Decrease & BDPCM, CCALF, LMCS \\
  \hline \hline
 & LB \\
\hline \hline
Major Increase & DQ, GPM, MMVD \\
\hdashline
Minor Increase & AMVR, BCW, CCLM, CST, ISP, JCCR, MIP, MRL \\
\hline 
Major Decrease & AFFINE, ALF, DBF, LMCS, SAO, SBTMVP \\
\hdashline
\multirow{2}{*}{Minor Decrease} & BDPCM, CCALF, CIIP, IBC \\
	& LFNST, MTS, PROF, SBT  \\

\hline \hline 
  & RA  \\
\hline \hline
Major Increase  & BCW, DQ, GPM \\
\hdashline  
\multirow{2}{*}{Minor Increase} & AMVR, BDPCM, CST, IBC, JCCR, MMVD, MRL \\
& SMVD  \\
\hline
Major Decrease &  AFFINE, ALF, BDOF, DBF, DMVR, SAO \\
\hdashline
\multirow{2}{*}{Minor Decrease} &  CCALF, CCLM, CIIP, ISP, LFNST, LMCS, MIP \\
            & MTS, PROF, SBT, SBTMVP \\
\end{tabular}
\end{center}
\vspace*{-0.5cm}
\end{table}

In Table~\ref{tab:CodingToolDescription}, the assignment to each category is shown for all coding tools. A major energy efficiency increase is measured for the coding tools: CST, DQ, JCCR, and MRL. For the coding tools ALF, DBF, IBC, ISP, LFNST, and SAO, the energy efficiency is decreased significantly. Therefore, the decoder energy demand of VVC can be reduced if in-loop filters such as ALF, DBF, and SAO are not used by the encoder.

\subsubsection{Lowdelay B}
For the coding tools DQ, GPM, and MMVD, we observe a major energy efficiency increase. In Figure~\ref{fig:Training}~(b), these tools have a positive BDDE-PSNR value of more than \pro{1}. For GPM, the decoder's energy demand is increased by \pro{4.27} if the coding tool is disabled, and the bit rate is increased by \pro{2.24}.

For the coding tools AFFINE, ALF, DBF, LMCS, SAO, and SBTMVP, the energy efficiency is decreased significantly.
In particular, DBF has a significant impact on the energy efficiency with a BDDE-PSNR value \pro{-20.04}. Simultaneously, the bit rate is slightly increased with a BDR-PSNR of \pro{0.86} (c.f. Figure~\ref{fig:Training}~(b)). Furthermore, for AFFINE, we determine a BDDE-PSNR value of \pro{-5.11} and a BDR-PSNR value of \pro{1.54}. For ALF, the BDDE-PSNR value is \pro{-6.13} and the BDR-PSNR value \pro{3.99}. Therefore, ALF can significantly improve the compression efficiency by spending a significant amount of computational complexity.

For the EE profile, we use a profile that has a BDDE-PSNR value of \pro{-39.88} and a BDR-PSNR value of \pro{10.98}, which is a profile from the second iteration of the DSE training. For the EBE profile, we use another profile from the second iteration that has a BDDE-PSNR value of \pro{-32.70} and a BDR-PSNR value of \pro{4.69}. This profile was selected based on the refinement similar to AI.

\subsubsection{Randomaccess}

For RA coding, we determine that three coding tools (BCW, DQ, and GPM) have a major increase of the energy efficiency. The highest increase in decoding energy can be observed for the coding tool GPM (BDDE: \pro{2.91} and BDR: \pro{1.84}) if the tool is disabled. This can be explained by the functionality of the coding tool, which is the prediction of motion with blocks that have a triangular or trapezoid shape~\cite{Blaeser2019}. Thereby, it is not necessary to have further splits of a CU to represent complex motion, which results in higher block sizes if GPM is used. For such blocks, GPM avoids subpartitioning into multiple, small rectangular blocks such that the overall share of large block sizes increases. The lower share of blocks with less pixels is observed by the bit stream feature analyzer proposed in~\cite{Kraenzler2020ICIP}, where for the CTC profile, \pro{36.70} pixels are predicted with a block size of less than 512 pixels and for the profile with disabled GPM, the share is \pro{37.80}.  In total, GPM is used for the CTC profile for a share of \pro{5.66} pixels. As a consequence, the complexity decrease related to a lower number of small blocks overcompensates the complexity increase due to GPM.

Finally, for the coding tools AFFINE, ALF, BDOF, DBF, DMVR, and SAO, the energy efficiency is decreased significantly when disabled. Again, DBF has the highest energy demand reduction (BDDE-PSNR: \pro{-12.84}). For ALF, the energy demand is reduced by \pro{-5.57} and the bit rate is increased by \pro{3.63} if the tool is disabled. Correspondingly, for the coding tool DMVR, the bit rate is increased slightly (BDR-PSNR: \pro{0.47}) and the energy demand is reduced by \pro{-4.61}.

For the EE profile, we choose the profile that has the minimal BDDE-PSNR value of \pro{-44.13} and a BDR-PSNR value of \pro{17.08}. The usage of the corresponding coding tool is shown in Table~\ref{tab:EnergyEfficientConfigs}. For the EBE profile, we choose another profile with a BDDE-PSNR value of \pro{-36.33} and a BDR-PSNR value of \pro{9.99} based on the earlier described refinement.

\begin{table}[t!]
\def\arraystretch{1.05}
\caption{Proposed EE and EBE profiles for all CTC configurations as derived by the DSE.}
\label{tab:EnergyEfficientConfigs}
\begin{center}
\begin{tabular}{ l | c  c  | c   c | c c} 
            &  \multicolumn{2}{c|}{AI} &  \multicolumn{2}{c|}{LB} &  \multicolumn{2}{c}{RA} \\
  Tool      &  EE & EBE & EE & EBE & EE & EBE\\
\hline\hline
  CCLM   & $\xmark$ & $\cmark$ & $\cmark$ &  $\cmark$ & $\xmark$ & $\xmark$ \\
  ISP    & $\xmark$ & $\cmark$ & $\xmark$ &  $\cmark$ & $\xmark$ & $\xmark$ \\
  MIP    & $\xmark$ & $\cmark$ & $\cmark$ &  $\cmark$ & $\xmark$ & $\xmark$ \\
  MRL   & $\cmark$ & $\cmark$ & $\cmark$ &  $\cmark$ & $\cmark$ & $\cmark$  \\
 \hline\hline
  AFFINE & -        & -        & $\xmark$ &  $\xmark$ & $\xmark$ & $\xmark$ \\
  AMVR   & -        & -        & $\cmark$ &  $\cmark$ & $\cmark$ & $\cmark$ \\
  BCW    & -        & -        & $\cmark$ &  $\cmark$ & $\xmark$ & $\cmark$ \\
  BDOF   & -        & -        & -        &  -        & $\xmark$ & $\xmark$ \\
  CIIP   & -        & -        & $\xmark$ &  $\xmark$ & $\xmark$ & $\xmark$ \\
  DMVR   & -        & -        & -        &  -        & $\xmark$ & $\xmark$ \\
  GPM    & -        & -        & $\cmark$ &  $\cmark$ & $\cmark$ & $\cmark$ \\
  MMVD   & -        & -        & $\cmark$ &  $\cmark$ & $\cmark$ & $\cmark$ \\
  PROF   & -        & -        & $\xmark$ &  $\xmark$ & $\xmark$ & $\xmark$ \\ 
  SBTMVP & -        & -        & $\xmark$ &  $\xmark$ & $\xmark$ & $\xmark$ \\
  SMVD   & -        & -        & -        &  -        & $\xmark$ & $\xmark$ \\
\hline\hline
  DQ     & $\cmark$ & $\cmark$ & $\cmark$ &  $\cmark$ & $\cmark$ & $\cmark$ \\
  JCCR   & $\cmark$ & $\cmark$ & $\cmark$ &  $\cmark$ & $\cmark$ & $\cmark$ \\
  LFNST  & $\xmark$ & $\cmark$ & $\xmark$ &  $\xmark$ & $\xmark$ & $\xmark$ \\
  MTS    & $\xmark$ & $\cmark$ & $\xmark$ &  $\xmark$ & $\xmark$ & $\xmark$ \\
  SBT    & -        & -        & $\xmark$ &  $\xmark$ & $\xmark$ & $\xmark$ \\
\hline\hline 
  ALF    & $\xmark$ & $\xmark$ & $\xmark$ &  $\cmark$ & $\xmark$ & $\cmark$ \\
  CCALF  & $\xmark$ & $\xmark$ & $\xmark$ &  $\xmark$ & $\xmark$ & $\xmark$ \\
  DBF    & $\xmark$ & $\xmark$ & $\xmark$ &  $\xmark$ & $\xmark$ & $\xmark$\\
  LMCS   & $\xmark$ & $\xmark$ & $\xmark$ &  $\xmark$ & $\xmark$ & $\xmark$ \\
  SAO    & $\xmark$ & $\xmark$ & $\xmark$ &  $\xmark$ & $\xmark$ & $\xmark$ \\
\hline\hline
 BDPCM   & $\xmark$ & $\xmark$ & $\xmark$ &  $\xmark$ & $\cmark$ & $\cmark$ \\
  IBC    & $\xmark$ & $\xmark$ & $\xmark$ &  $\xmark$ & $\cmark$ & $\cmark$ \\
  CST    & $\cmark$ & $\cmark$ & $\cmark$ &  $\cmark$ & $\cmark$ & $\cmark$ \\
  \hline \hline
\end{tabular}
\end{center}
\vspace*{-0.5cm}
\end{table}

\subsection{Validation}
\label{subsec:Validation}

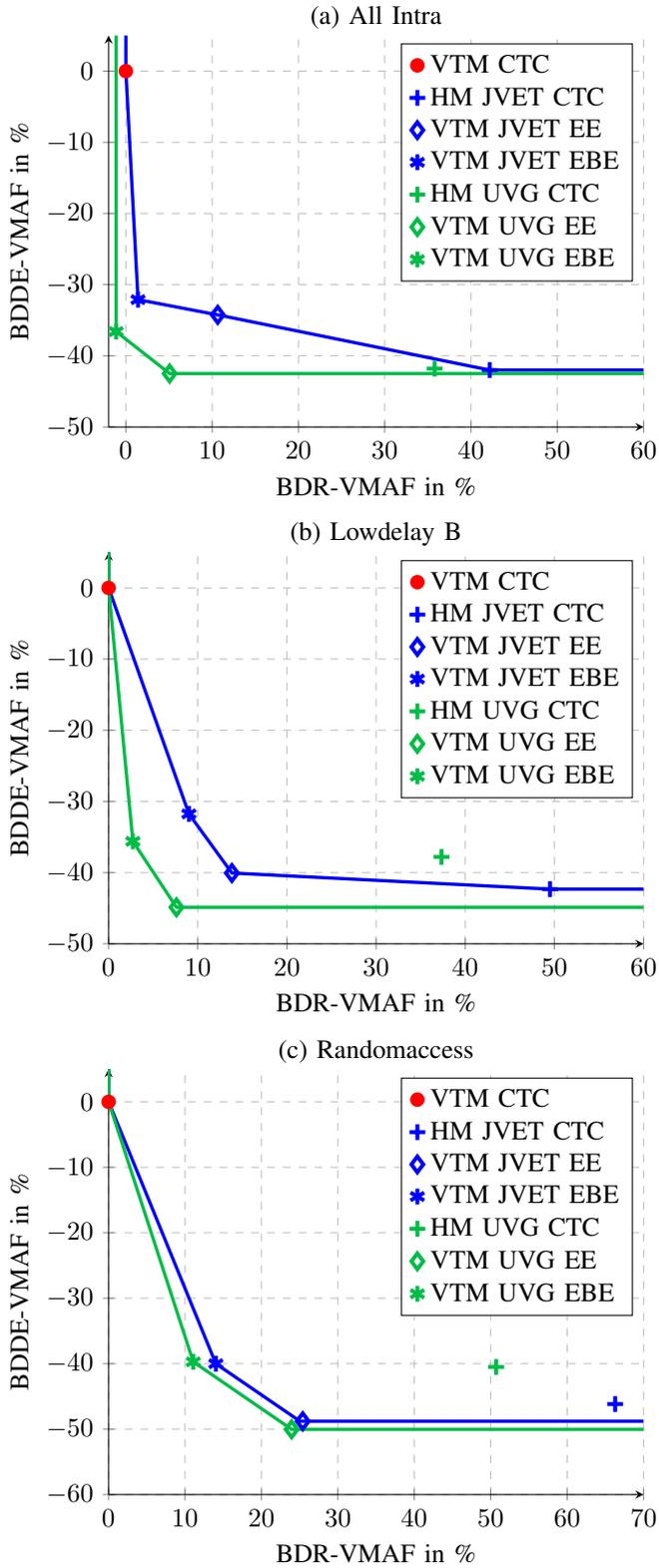
\begin{figure}[!t]
\definecolor{Iter1}{HTML}{40D4EB}
\definecolor{Iter2}{HTML}{FF7100}
\definecolor{Iter3}{HTML}{FFF100}
\definecolor{EE}{HTML}{00BD47}
\begin{center}
\begin{tikzpicture}
\begin{groupplot}[
     group style={group name=my plots,group size= 1 by 3,vertical sep =2.00cm},
     title style={yshift=-0.25cm},
     height = 10cm,
     width = \textwidth,
    ]
    
    \nextgroupplot[
	width=0.48\textwidth,
	height = 0.375\textwidth,
    xlabel={BDR-VMAF in \%},
    ylabel={BDDE-VMAF in \%},
    title = {(a) All Intra},
    legend cell align={left},
    xmin=-2, xmax=60,
    ymin=-50, ymax=5,
        axis lines = left,
    ymajorgrids=true,
    xmajorgrids=true,
    xminorgrids=true,
yminorgrids=true,
    grid style=dashed,
        xtick={0,10,20,30,40,50,60},
    ytick={0,-10,-20,-30,-40,-50},
]

\addplot[only marks,
   color=red,
    mark=*,
    line width=1.5pt,
        mark size = 2pt,
    ]
    coordinates {
    (0,0)
    };
    \addlegendentry{VTM CTC}

\addplot[only marks,
   color=blue,
    mark=+,
    line width=1.5pt,
    mark size = 3pt,
    ]
    coordinates {
    (42.17,-42.00)
    };
    \addlegendentry{HM JVET CTC}

\addplot[only marks,
   color=blue,
    mark=diamond,
    line width=1.5pt,
    mark size = 3pt,
    ]
    coordinates {
 (10.65,-34.25)
    };
    \addlegendentry{VTM JVET EE}

\addplot[only marks,
   color=blue,
    mark=asterisk,
    line width=1.5pt,
    mark size = 3pt,
    ]
    coordinates {
   (1.41,-32.11)
    };
    \addlegendentry{VTM JVET EBE}
    
\addplot[only marks,
   color=EE,
    mark=+,
    line width=1.5pt,
    mark size = 3pt,
    ]
    coordinates {
    (35.79,-41.80)
    };
    \addlegendentry{HM UVG CTC}    
    
    \addplot[only marks,
   color=EE,
    mark=diamond,
    line width=1.5pt,
    mark size = 3pt,
    ]
    coordinates {
 (5.07,-42.49)
    };
    \addlegendentry{VTM UVG EE}

    \addplot[only marks,
   color=EE,
    mark=asterisk,
    line width=1.5pt,
    mark size = 3pt,
    ]
    coordinates {
 (-1.13,-36.59)
    };
    \addlegendentry{VTM UVG EBE}

\draw [ 
   color=blue,
    mark=none,
    mark size=2pt,
    line width=1.25pt
    ] plot coordinates {   
  (0,5)
  (0,0)
  (1.41,-32.11)
  (10.65,-34.25)
  (42.17,-42.00)
  (60.0,-42.00)
  };

\draw [ 
   color=EE,
    mark=none,
    mark size=2pt,
    line width=1.25pt
    ] plot coordinates {   (-1.13,5) (-1.13,0) (-1.13,-36.59) (5.07,-42.49)(60.0,-42.49)  };

\nextgroupplot[
	width=0.48\textwidth,
	height = 0.375\textwidth,
    xlabel={BDR-VMAF in \%},
    ylabel={BDDE-VMAF in \%},
    title = {(b) Lowdelay B},
    xmin=0, xmax=60,
        xtick={0,10,20,30,40,50,60},
    ytick={0,-10,-20,-30,-40,-50},
legend cell align={left},
    ymin=-50, ymax=5,
        axis lines = left,
    ymajorgrids=true,
    xmajorgrids=true,
    xminorgrids=true,
yminorgrids=true,
    grid=both,
    grid style=dashed,
    yshift=0.33cm,
]

\addplot[only marks,
   color=red,
    mark=*,
    line width=1.5pt,
        mark size = 2pt,
    ]
    coordinates {
    (0,0)
    };
    \addlegendentry{VTM CTC}

\addplot[only marks,
   color=blue,
    mark=+,
    line width=1.5pt,
    mark size = 3pt,
    ]
    coordinates {
    (49.51,-42.32)
    };
    \addlegendentry{HM JVET CTC}

\addplot[only marks,
   color=blue,
    mark=diamond,
    line width=1.5pt,
    mark size = 3pt,
    ]
    coordinates {
 (13.83,-40.06)
    };
    \addlegendentry{VTM JVET EE}

\addplot[only marks,
   color=blue,
    mark=asterisk,
    line width=1.5pt,
    mark size = 3pt,
    ]
    coordinates {
   (9.01,-31.75)
    };
    \addlegendentry{VTM JVET EBE}
    
\addplot[only marks,
   color=EE,
    mark=+,
    line width=1.5pt,
    mark size = 3pt,
    ]
    coordinates {
    (37.34,-37.80)
    };
    \addlegendentry{HM UVG CTC}    
    
    \addplot[only marks,
   color=EE,
    mark=diamond,
    line width=1.5pt,
    mark size = 3pt,
    ]
    coordinates {
 (7.61,-44.87)
    };
    \addlegendentry{VTM UVG EE}

    \addplot[only marks,
   color=EE,
    mark=asterisk,
    line width=1.5pt,
    mark size = 3pt,
    ]
    coordinates {
 (2.72,-35.61)
    };
    \addlegendentry{VTM UVG EBE}

\draw [ 
   color=blue,
    mark=none,
    mark size=2pt,
    line width=1.25pt
    ] plot coordinates {   
  (0,5)
  (0,0)
  (9.01,-31.75)
  (13.83,-40.06)
  (49.51,-42.32)
  (60.0,-42.32)
  };

\draw [ 
   color=EE,
    mark=none,
    mark size=2pt,
    line width=1.25pt
    ] plot coordinates {   (0,5)
  (0,0)
  (2.72,-35.61)
  (7.61,-44.87)
  (60.0,-44.87)
  };

\nextgroupplot[
	width=0.48\textwidth,
	height = 0.4\textwidth,
    xlabel={BDR-VMAF in \%},
    ylabel={BDDE-VMAF in \%},
    title = {(c) Randomaccess},
    xmin=0, xmax=70,
legend cell align={left},
    ymin=-60, ymax=5,
        axis lines = left,
            xtick={0,10,20,30,40,50,60,70},
    ytick={0,-10,-20,-30,-40,-50,-60},
    ymajorgrids=true,
    xmajorgrids=true,
    xminorgrids=true,
yminorgrids=true,
    grid style=dashed,
        yshift=0.33cm,
]

\addplot[only marks,
   color=red,
    mark=*,
    line width=1.5pt,
        mark size = 2pt,
    ]
    coordinates {
    (0,0)
    };
    \addlegendentry{VTM CTC}

\addplot[only marks,
   color=blue,
    mark=+,
    line width=1.5pt,
    mark size = 3pt,
    ]
    coordinates {
    (66.28,-46.17)
    };
    \addlegendentry{HM JVET CTC}

\addplot[only marks,
   color=blue,
    mark=diamond,
    line width=1.5pt,
    mark size = 3pt,
    ]
    coordinates {
 (25.40,-48.79)
    };
    \addlegendentry{VTM JVET EE}

\addplot[only marks,
   color=blue,
    mark=asterisk,
    line width=1.5pt,
    mark size = 3pt,
    ]
    coordinates {
   (14.04,-40.01)
    };
    \addlegendentry{VTM JVET EBE}
    
\addplot[only marks,
   color=EE,
    mark=+,
    line width=1.5pt,
    mark size = 3pt,
    ]
    coordinates {
    (50.70,-40.51)
    };
    \addlegendentry{HM UVG CTC}    
    
    \addplot[only marks,
   color=EE,
    mark=diamond,
    line width=1.5pt,
    mark size = 3pt,
    ]
    coordinates {
 (23.95,-50.03)
    };
    \addlegendentry{VTM UVG EE}

    \addplot[only marks,
   color=EE,
    mark=asterisk,
    line width=1.5pt,
    mark size = 3pt,
    ]
    coordinates {
 (11.08,-39.72)
    };
    \addlegendentry{VTM UVG EBE}

\draw [ 
   color=blue,
    mark=none,
    mark size=2pt,
    line width=1.25pt
    ] plot coordinates {   
  (0,5)
  (0,0)
  (14.04,-40.01)
  (25.04,-48.79)
  (70.0,-48.79)
  };

\draw [ 
   color=EE,
    mark=none,
    mark size=2pt,
    line width=1.25pt
    ] plot coordinates {   (0,5)
  (0,0)
  (11.08,-39.72)
  (23.95,-50.03)
  (70.0,-50.03)
  };

\end{groupplot}
	
\end{tikzpicture}
\end{center}

\vspace*{-0.5cm}
\caption{Results of the proposed EE and EBE coding tool profiles for both sequence sets. In addition, the results of the VTM and HM CTC profile are shown for each coding configuration. The horizontal axis shows the additional bit rate measured in BDR-VMAF in relation to the VTM CTC profile, and the vertical axis shows the energy demand reduction measured in BDDE-VMAF, respectively. In (a), the results of all intra coding are shown, in (b) for lowdelay B, and in (c) for randomaccess. For each sequence set, a Pareto curve is depicted for each coding configuration.}

\label{fig:ValidationPlots}

\end{figure}

In the following, we will evaluate the proposed coding tool profiles of Table~\ref{tab:EnergyEfficientConfigs} on both video sets introduced in Section~\ref{subsec:testset}. Therefore, we will at first take a look at the impact of the EE and EBE profile in terms of bit rate and decoding energy demand compared to the CTC profile with the VTM decoder. The results of this evaluation are shown in Figure~\ref{fig:ValidationPlots} and in Table~\ref{tab:ValidationResults}. For all BD metrics, we use the CTC profile with the VTM decoder as a reference. Finally, we will compare both profiles for the VVdeC decoder in Table~\ref{tab:VVdeCValidationResults}.

\begin{table*}[t!]
\caption{Evaluation of the two proposed coding tool profiles in comparison to the CTC coding tool profiles, which are encoded with VTM. Furthermore, the VTM coded bit streams are compared to the CTC HM coded bit streams. For all, the BDR and BDDE is calculated with both PSNR and VMAF as a visual quality metric and the VTM CTC profile as a reference.}
\label{tab:ValidationResults}
\begin{center}
\begin{tabular}{c || c : c  | c : c  || c : c  | c  : c  || c : c | c : c} 
  & \multicolumn{2}{c|}{BDR in $\%$} & \multicolumn{2}{c||}{BDDE in $\%$} & \multicolumn{2}{c|}{BDR in $\%$} & \multicolumn{2}{c||}{BDDE in $\%$} &  \multicolumn{2}{c|}{BDR in $\%$} & \multicolumn{2}{c}{BDDE in $\%$}\\ 
 &  PSNR & VMAF & PSNR & VMAF & PSNR & VMAF & PSNR & VMAF & PSNR & VMAF & PSNR & VMAF   \\
 \hline\hline 
 & \multicolumn{4}{c||}{HM AI CTC vs. VTM CTC} & \multicolumn{4}{c||}{VTM AI EE vs. VTM CTC}
 & \multicolumn{4}{c}{VTM AI EBE vs. VTM CTC} \\
\hline
    A1 & 43.79 & 50.08 & -40.57 & -39.24 & 32.80 & 14.65 & -34.63 & -39.52 & 8.98 & 3.62 & -34.22 & -37.60 \\ 
     A2 & 40.01 & 47.50 & -44.30 & -42.54 & 19.85 & 7.35 & -34.06 & -40.13 & 9.73 & 2.47 & -30.75 & -35.75 \\ 
     B & 29.93 & 34.05 & -45.15 & -42.32 & 16.81 & 6.93 & -33.77 & -37.85 & 6.94 & 2.05 & -29.83 & -33.71 \\ 
     C & 29.72 & 31.00 & -48.11 & -45.73 & 14.18 & 4.43 & -27.05 & -31.70 & 4.77 & -0.35 & -25.72 & -29.67 \\ 
     D & 21.38 & 25.60 & -48.89 & -46.60 & 10.27 & 3.10 & -25.31 & -28.99 & 3.23 & -0.76 & -23.75 & -27.17 \\ 
     E & 34.56 & 35.78 & -40.41 & -39.22 & 17.23 & 5.34 & -32.60 & -35.55 & 7.45 & 0.18 & -31.95 & -34.81 \\ 
     F & 73.29 & 74.91 & -38.85 & -37.02 & 37.32 & 32.53 & -25.23 & -28.23 & 6.66 & 3.04 & -25.96 & -28.62 \\ 
     
     \hdashline
      JVET  & \textbf{38.55} & \textbf{42.17} & \textbf{-44.04} & \textbf{-42.00} & \textbf{20.80} & \textbf{10.65} & \textbf{-30.12} & \textbf{-34.25} & \textbf{6.61} & \textbf{1.41} & \textbf{-28.53} & \textbf{-32.11} \\ 
      \hline 
     UVG1 & 30.60 & 34.64 & -40.93 & -39.58 & 18.65 & 1.85 & -38.94 & -42.63 & 5.18 & -3.65 & -33.37 & -36.91 \\ 
     UVG2 & 40.01 & 37.86 & -43.41 & -42.79 & 25.36 & 6.42 & -38.67 & -42.56 & 9.59 & -0.24 & -34.91 & -38.43 \\ 
     UVG3 & 35.89 & 34.75 & -44.86 & -43.22 & 21.23 & 7.20 & -39.48 & -42.23 & 5.48 & 0.71 & -31.95 & -34.11 \\ 
     \hdashline
     UVG & \textbf{35.48} & \textbf{35.79} & \textbf{-42.99} & \textbf{-41.80} & \textbf{21.77} & \textbf{5.07} & \textbf{-39.01} & \textbf{-42.49} & \textbf{6.80} & \textbf{-1.13} & \textbf{-33.47} & \textbf{-36.59} \\ 
  \hline\hline & \multicolumn{4}{c||}{HM LB CTC vs. VTM CTC} & \multicolumn{4}{c||}{VTM LB EE vs. VTM CTC} 
  & \multicolumn{4}{c}{VTM LB EBE vs. VTM CTC}\\
\hline
B & 47.07 & 48.13 & -42.66 & -42.11 & 22.86 & 15.60 & -43.59 & -46.20 & 13.22 & 8.22 & -34.74 & -36.47 \\ 
C & 40.27 & 39.95 & -45.15 & -45.30 & 13.51 & 7.82 & -35.97 & -39.81 & 6.18 & 5.18 & -29.42 & -31.29 \\ 
D & 32.82 & 35.22 & -47.63 & -47.90 & 14.68 & 11.39 & -36.11 & -40.56 & 9.27 & 6.90 & -29.97 & -33.20 \\ 
E & 52.25 & 45.27 & -41.19 & -41.62 & 17.26 & 8.16 & -39.79 & -41.01 & 9.65 & 6.32 & -33.22 & -34.51 \\ 
F & 80.54 & 78.28 & -34.34 & -34.55 & 28.41 & 24.34 & -29.34 & -31.42 & 18.13 & 17.93 & -21.82 & -22.80 \\ 
         \hdashline
JVET & \textbf{50.33} & \textbf{49.51} & \textbf{-42.27} & \textbf{-42.32} & \textbf{19.62} & \textbf{13.83} & \textbf{-37.15} & \textbf{-40.06} & \textbf{11.47} & \textbf{9.01} & \textbf{-29.91} & \textbf{-31.75} \\ 
      \hline 
     UVG3 & \textbf{40.96} & \textbf{37.34} & \textbf{-37.87} & \textbf{-37.80} & \textbf{13.66} & \textbf{7.61} & \textbf{-43.48} & \textbf{-44.87} & \textbf{5.68} & \textbf{2.72} & \textbf{-34.63} & \textbf{-35.61} \\  
  \hline\hline & \multicolumn{4}{c||}{HM RA CTC vs. VTM CTC} & \multicolumn{4}{c||}{VTM RA EE vs. VTM CTC} 
  & \multicolumn{4}{c}{VTM RA EBE vs. VTM CTC}\\
\hline
A1 & 64.68 & 73.57 & -44.92 & -45.11 & 36.96 & 29.22 & -52.84 & -54.67 & 24.45 & 19.06 & -41.24 & -42.70 \\ 
     A2 & 75.82 & 97.80 & -50.08 & -49.14 & 41.10 & 47.12 & -55.85 & -56.57 & 25.26 & 20.85 & -45.72 & -46.99 \\ 
     B & 59.49 & 66.43 & -45.69 & -45.82 & 32.67 & 30.58 & -51.35 & -53.40 & 20.35 & 14.41 & -41.59 & -43.83 \\ 
     C & 43.89 & 44.43 & -46.02 & -46.39 & 21.23 & 16.37 & -42.32 & -45.84 & 13.51 & 10.60 & -34.73 & -37.41 \\ 
     D & 38.58 & 45.01 & -47.13 & -47.15 & 20.14 & 21.06 & -41.78 & -44.97 & 13.13 & 10.20 & -35.17 & -38.35 \\ 
       F & 78.39 & 80.12 & -43.87 & -43.99 & 21.70 & 13.13 & -37.34 & -39.54 & 14.83 & 11.98 & -30.79 & -32.23 \\ 
     
     \hdashline
      JVET & \textbf{59.23} & \textbf{66.28} & \textbf{-46.15} & \textbf{-46.17} & \textbf{28.25} & \textbf{25.40} & \textbf{-46.46} & \textbf{-48.79} & \textbf{18.12} & \textbf{14.04} & \textbf{-37.89} & \textbf{-40.01} \\ 
      \hline 
     UVG1 & 49.69 & 43.73 & -38.63 & -40.36 & 24.96 & 20.67 & -50.08 & -51.22 & 12.37 & 11.43 & -37.78 & -39.24 \\ 
     UVG2 & 64.61 & 65.07 & -37.80 & -38.12 & 34.65 & 31.54 & -48.88 & -48.10 & 19.60 & 12.16 & -38.34 & -38.61 \\ 
     UVG3 & 46.06 & 42.25 & -43.19 & -43.41 & 23.74 & 19.04 & -49.78 & -50.87 & 13.70 & 9.44 & -40.24 & -41.53 \\ 
    \hdashline
     UVG & \textbf{53.77} & \textbf{50.70} & \textbf{-39.73} & \textbf{-40.51} & \textbf{27.96} & \textbf{23.95} & \textbf{-49.57} & \textbf{-50.03} & \textbf{15.29} & \textbf{11.08} & \textbf{-38.72} & \textbf{-39.72} \\ 
  \hline\hline 
\end{tabular}
\end{center}
\end{table*}

\subsubsection{All Intra}
 
 For the EE profile, we determine that we have a BDR-VMAF of \pro{10.65} for the JVET set and of \pro{5.07} for the UVG set, which is significantly less than for HEVC with over \pro{30} (c.f. Tab.~\ref{tab:ValidationResults}). The energy demand is reduced by \pro{-34.25} (BDDE-VMAF) on average for the JVET set and by \pro{-42.49} for the UVG set. Furthermore, we observe from the results of Table~\ref{tab:ValidationResults} that the energy demand of the EE profile is comparable to the HM decoder. However, the bit rate increase is significantly lower on average than for HEVC.

 With the EBE profile, the additional bit rate for the JVET set is reduced to \pro{1.41} (BDDR-VMAF). Therefore, we have a similar compression performance with respect to the CTC profile of VVC. For the UVG set, which has a BDR-VMAF value of \pro{-1.13}, the bit rate with respect to the CTC profile is even improved, if the same VMAF quality score is targeted. For the energy efficiency, the EBE profile has a BDDE-VMAF value of \pro{-32.11} for the JVET set and of \pro{-36.59} for the UVG set.

In Figure~\ref{fig:ValidationPlots}, the compression and energy efficiency for both proposed coding tool profiles and both sequence sets are illustrated. The horizontal axis in each plot shows the BDR-VMAF value and the vertical axis the BDDE-VMAF value. In the plots, a lower BDDE-VMAF value corresponds to a higher energy efficiency and a lower BDR-VMAF value correspond to a higher compression efficiency, which means that a profile at the bottom left corner is desirable. In the figure, the diamond-shaped marker corresponds to the EE profile, the asterisk-shaped markers to the EBE profile, the plus-shaped markers to the HM CTC profile, and the circle-shaped markers to the VTM CTC profile, which is the reference for the calculation of the BD metrics. The blue markers correspond to the JVET set and the green markers to the UVG set.

For both Pareto curves, we determine that the EE and the EBE profile in Figure~\ref{fig:ValidationPlots}~(a) reduce the bit rate significantly in relation to the HM CTC profile. For the UVG set, the energy efficiency of the EE profile is slightly higher than the HM CTC profile and the compression efficiency is higher than for the VTM CTC profile.

\subsubsection{Lowdelay B}
Similar to AI coding, the energy demand of LB can be reduced by about \pro{42} (c.f. Table~\ref{tab:ValidationResults}) for the JVET set and by \pro{38} for the UVG set if the bit streams are encoded with HEVC instead of VVC. As a downside of HEVC, the bit rate is increased by approximately~\pro{50} and \pro{40}, respectively.

For the EE profile, we have a BDR-PSNR value of \pro{19.62} for the JVET set and of \pro{13.66} for the UVG set. According to Table~\ref{tab:ValidationResults}, the energy demand is reduced by \pro{-40.06} (BDDE-VMAF) with a bit rate increase of \pro{13.83} (BDR-VMAF) for the JVET set. For the UVG set, the energy reduction is even higher with \pro{-44.87} and a bit rate increase of \pro{7.61}. 

For the EBE profile of LB coding, the BDR-VMAF is \pro{9.01} and the BDDE-VMAF \pro{-31.75} for the JVET set. For the UVG set, the bit rate increase is \pro{2.72} in terms of BDR-VMAF and the energy demand is reduced by \pro{-35.61}~(BDDE-VMAF). 

In Figure~\ref{fig:ValidationPlots}~(b), we determine that the energy efficiency of the EE profile is significantly higher than of the HM CTC profile for the UVG set. For the EBE profile, the energy efficiency is slightly less than the HM CTC profile for both video sequence sets. 

\subsubsection{Randomaccess}

Strikingly, the BDDE-PSNR and BDDE-VMAF values of HM are outperformed by the RA EE profile for both sequence sets. This is shown in Figure~\ref{fig:ValidationPlots}~(c) by both Pareto curves, which are below the markers of the HM CTC profile. Furthermore, the energy efficiency of the EBE profile for the UVG set is similar to the HM CTC profile.

For the EE profile, we measured a BDDE-VMAF value of \pro{-48.79} for the JVET set and of \pro{-50.03} for the UVG set, which means that the energy demand of those sets can be halved in relation to the CTC coded bit streams. In particular, the bit streams of class A1, A2, UVG1, UVG3, and B have a BDDE-VMAF value of under \pro{-50} (c.f. Table~\ref{tab:ValidationResults}). Therefore, we observe the highest energy demand reduction for sequences with a very high resolution. From Figure~\ref{fig:ValidationPlots}, we observe that both the energy and compression efficiency are improved by the EE profile in relation to HEVC. For the EBE profile for RA coding, we show that the energy demand can be reduced by approximately \pro{-40}. In terms of bit rate, the BDR-VMAF value for the JVET set is \pro{14.04} and for the UVG set \pro{11.08}.

For class F, which comprises screen-captured or computer-generated content, we can determine that both for RA and LB the two proposed coding tool profiles have the lowest energy savings compared to other classes. This can be explained by the different characteristics of the sequences. Therefore, several coding tools have been introduced in VVC, such as palette mode, adaptive color transform, and transform skip residual coding, to improve the compression efficiency of screen content in addition to IBC and BDPCM~\cite{Bross2021a}. In the future, the influence of these types of sequences can be studied in more detail.

In general, we observe from our evaluation that both BD metrics with VMAF are improved by a higher degree than with PSNR, which is shown by a lower BD value. Since VMAF has a higher correlation to the subjective visual quality than PSNR, this is desirable.

Besides the decoder, we also evaluated the influence of both proposed profiles on the encoding time. For the sequence MarketPlace, we encoded the first 65 frames with RA coding. To evaluate the encoding time demand, we use the metric Bj{\o}ntegaard-Delta encoding time (BDET) with PSNR as objective quality metric. As a reference for the calculation of BDET-PSNR and BDR-PSNR, we use the VTM CTC profile. For the EBE profile, we measure a BDET-PSNR of \pro{-42.43} and BDR-PSNR of \pro{23.49}. For the EE profile, the BDET-PSNR value is \pro{-49.58} and the BDR-PSNR value is \pro{34.78}, which means that the encoding time can be reduced by approximately \pro{50}.

\begin{table}[!t]
\def\arraystretch{1.05}
\caption{Evaluation of the energy efficiency of the proposed coding tool profiles and the openHEVC decoder in relation to the CTC encoded VTM bit streams decoded with VVdeC on x86-CPU.}
\label{tab:VVdeCValidationResults}
\begin{center}
\begin{tabular}{c || c : c | c : c | c : c } 
& \multicolumn{2}{c|}{openHEVC }
& \multicolumn{2}{c|}{VVdeC EE}
& \multicolumn{2}{c}{VVdeC EBE}  \\
\hline 
  & \multicolumn{2}{c|}{BDDE in $\%$} 
  & \multicolumn{2}{c|}{BDDE in $\%$} 
  & \multicolumn{2}{c}{BDDE in $\%$}\\ 
 &  PSNR & VMAF 
 &  PSNR & VMAF 
 &  PSNR & VMAF \\
 \hline
  & \multicolumn{6}{c}{All Intra} \\
 \hline 
      JVET  & -63.16 & -64.42 & -33.26 & -38.37 & -28.18 & -32.58 \\
      \hdashline
      UVG  & -62.78 & -64.95 & -42.15 & -47.55 & -36.82 & -41.65 \\

 \hline
  & \multicolumn{6}{c}{Lowdelay B} \\
 \hline 
       JVET  & -42.40 & -43.78 & -33.83 & -34.88 & -21.83 & -22.86 \\ 
       \hdashline
       UVG & -22.28 & -22.52 & -31.26 & -30.44 & -18.19 & -17.80 \\ 
      
 \hline
  & \multicolumn{ 6}{c}{Randomaccess} \\
 \hline 
      JVET  & -45.49 & -45.87 & -35.48 & -37.07 & -24.12 & -25.53 \\  
     \hdashline
     UVG  & -31.27 & -31.22 & -38.95 & -40.17 & -22.71 & -23.63 \\ 
           \hline
\end{tabular}
\end{center}
\end{table}

\subsubsection{Analysis of Proposed Profiles for VVdeC}
\label{subsub:vvdec}
In the following, we will evaluate the energy efficiency of VVdeC on the x86-CPU. In Table~\ref{tab:VVdeCValidationResults}, we use the CTC profile with the VVdeC decoder as a reference. Since the bit rate is not affected by the usage of VVdeC, we only evaluate BDDE-PSNR and BDDE-VMAF. 

For AI coding, the energy demand of openHEVC is lower than for both proposed profiles. For the EE profile, we observe a BDDE-VMAF value of \pro{-38.37} for the JVET set and of \pro{-47.55} for the UVG set.
For LB coding, the EE profile has a higher decoding energy demand reduction than openHEVC for HD sequences (class B and UVG3). For RA coding, the energy demand of the UVG set can be reduced in relation to openHEVC. Therefore, the compression and energy efficiency of sequences with a high resolution is improved for RA and LB coding compared to HEVC. For the EBE profile, the energy demand is similar to openHEVC for RA and LB coding, but the compression efficiency is increased. Therefore, when using our proposed EE profile, VVdeC outperforms openHEVC both in terms of energy and compression efficiency for the UVG set.

For the ARM-CPU, we show the results of measurements on the Raspberry Pi in Table~\mbox{\ref{tab:VVdeCValidationResultsARM}}. The evaluation incorporates the coding configurations LB and RA, which are commonly used in streaming applications. Similar to Table~\mbox{\ref{tab:VVdeCValidationResults}}, the CTC coded bit streams decoded by VVdeC are the reference for the BDDE-PSNR and BDDE-VMAF calculations. Comparing openHEVC with VVdeC, for both coding configurations and video sequence sets, the energy savings are significantly higher on the ARM architecture, e.g, the JVET set has a BDDE-VMAF value of \mbox{\pro{-75.60}} for RA (for x86: \mbox{\pro{-45.97})}. This can be explained by the fact that VVdeC is mainly optimized for the x86-architecture\mbox{\cite{VVdec}}. For the EBE profile, the energy savings are similar to the x86-CPU. For RA coding, the BDDE-VMAF value is \mbox{\pro{-23.80}} for the JVET set and \mbox{\pro{-20.43}} for the UVG set (for x86: \mbox{\pro{-25.53}} and \mbox{\pro{-23.63}}, respectively). For the EE profile, the energy savings on the ARM-CPU are significantly higher for both coding configurations. For RA, the JVET set has a BDDE-value of \mbox{\pro{-59.66} and the UVG set of \mbox{\pro{-49.85}}}. Based on the findings in Section~\mbox{\ref{subsec:Training}}, where we concluded that the usage of the coding tool ALF leads to a major decrease in energy efficiency, we identify this coding tool as a reason for the higher energy savings on the ARM-CPU. The coding tool ALF is enabled for both EBE profiles and disabled for both EE profiles. Therefore, it can be assumed that the tool ALF is not yet fully optimized in VVdeC for the ARM architecture.

\section{Conclusion}
\label{sec:conclusion}
In this work, we propose a novel approach to optimize the energy demand of the VVC decoder with a greedy strategy based DSE. We show that the approach determines the optimal coding tool profile in terms of energy efficiency for a subset of features. Based on the algorithm, we derived multiple coding tool profiles that improve the energy efficiency of VVC significantly. We showed that the energy demand of the EE profiles can reduce the energy demand compared to the CTC coded VTM bit streams by over \pro{50}. By evaluating two decoder implementations of VVC and HEVC, we found that by using our proposed coding tool profiles for LB and RA encoding, we have a lower energy demand for VVC decoding than HEVC decoding.  Furthermore, we show that the complexity to determine such coding tool profiles can be reduced significantly. 

\begin{table}[!t]
\def\arraystretch{1.05}
\caption{Evaluation of the energy efficiency of the proposed coding tool profiles and the openHEVC decoder in relation to the CTC encoded VTM bit streams decoded with VVdeC on ARM-CPU.}

\label{tab:VVdeCValidationResultsARM}
\begin{center}
\begin{tabular}{c || c : c | c : c | c : c } 
& \multicolumn{2}{c|}{openHEVC }
& \multicolumn{2}{c|}{VVdeC EE}
& \multicolumn{2}{c}{VVdeC EBE}  \\
\hline 
  & \multicolumn{2}{c|}{BDDE in $\%$} 
  & \multicolumn{2}{c|}{BDDE in $\%$} 
  & \multicolumn{2}{c}{BDDE in $\%$}\\ 
 &  PSNR & VMAF 
 &  PSNR & VMAF 
 &  PSNR & VMAF \\
 \hline
  & \multicolumn{6}{c}{Lowdelay B} \\
 \hline 
       JVET  & -66.78 & -67.37 & -57.07 & -58.89 & -17.87 & -18.47\\ 
       \hdashline
       UVG & -67.91  & -67.26 & -54.21 & -54.65 & -15.00 & -15.60 \\ 
      
 \hline
  & \multicolumn{ 6}{c}{Randomaccess} \\
 \hline
       JVET  & -75.51 & -75.60 & -59.01 & -59.66 & -23.06 & -23.80 \\ 
       \hdashline
       UVG & -72.62 & -72.48 & -49.58  & -49.85 & -18.99 & -20.43 \\ 
           \hline
\end{tabular}
\end{center}
\vspace{-0.5cm}
\end{table}

Additionally, we analyzed the energy efficiency of two decoder implementations of VVC. For VTM, we determined that the energy demand increases by \pro{60} to \pro{80} compared to the HM decoder. For VVdeC, the energy demand is increased by \pro{30} up to \pro{180}, depending on the coding configuration and the resolution of the sequences.

From a tool sensitivity analysis, we determined that coding tools can be assigned to different categories, which evaluate the decoder's energy efficiency. For in-loop filter coding tools, we saw that the energy efficiency is decreased. However, for several coding tools (e.g., GPM), we saw that the energy efficiency of the decoder is improved.

In future work, we will study if the energy efficiency of the proposed profiles can be enhanced by DERDO~\cite{HerglotzHeindelKaup}. Also, block partitioning will be studied that might have further potential to reduce the decoder complexity. Furthermore, the influence of the proposed profiles on the encoding process will be analyzed in detail. 

The DSE algorithm can also be incorporated into the convex hull video encoding framework that is proposed in\mbox{\cite{Wu2021}}. For video streaming applications, adaptive bit rate (ABR) streaming is commonly used, which means that several encodings are processed with different quality levels and resolutions to optimize the video quality adaptively to the network connection speed. In\mbox{\cite{Wu2021}}, the authors propose a method that incorporates encodings with multiple QPs and resolutions to determine optimal encoding parameters for a given target quality or bit rate. However, with our proposed DSE, new coding tool profiles can be derived that optimize for the encoding complexity. 

Finally, the complexity of other state-of-the-art video codecs such as AV1 can be studied.

\ifCLASSOPTIONcaptionsoff
  \newpage
\fi

\bibliographystyle{IEEEtran}

\begin{IEEEbiography}[{\includegraphics[width=1in,height=1.25in,clip,keepaspectratio]{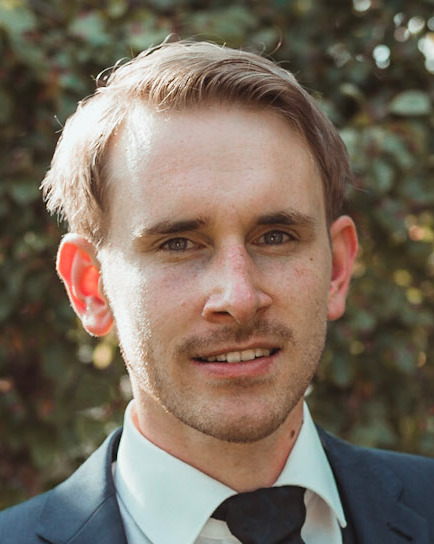}}]{Matthias Kr\"anzler} received the B.Sc. and M.Sc. degrees in electrical engineering and information technology from the Friedrich-Alexander University Erlangen-N\"urnberg (FAU), Germany, in 2017 and 2019, respectively. During his studies, he worked  as a student assistant on modeling and optimizing the energy demand of video decoders from 2015 to 2019. Since 2019, he has been a Research Scientist with the Chair of Multimedia Communications and Signal Processing, Friedrich-Alexander University Erlangen-N\"urnberg (FAU), Germany. His research interests include video coding and solutions for energy efficient video communication.
\end{IEEEbiography} 

\begin{IEEEbiography}[{\includegraphics[width=1in,height=1.25in,clip,keepaspectratio]{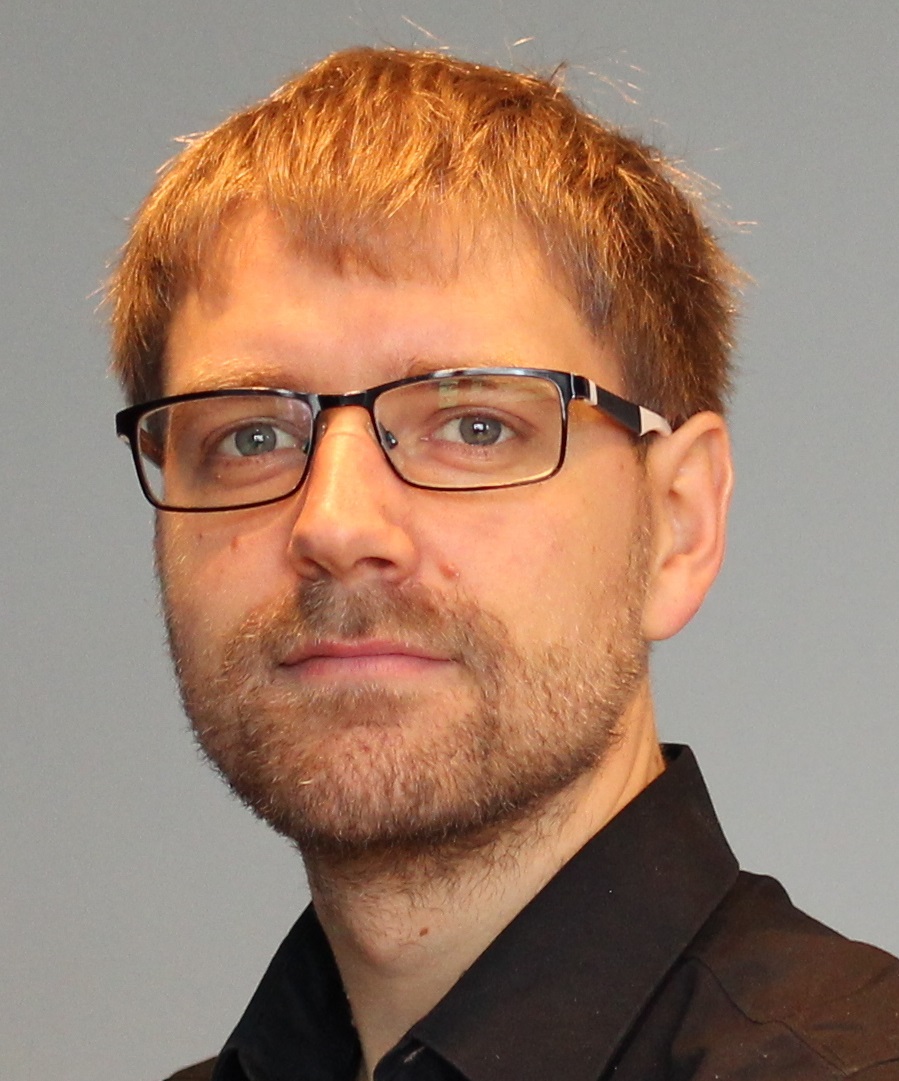}}] {Christian Herglotz} (Member, IEEE) received the Dipl.-Ing. in electrical engineering and information technology in 2011 and the Dipl.-Wirt. Ing. in business administration and economics in 2012, both from Rheinisch-Westfa\"lische Technische Hochschule (RWTH) Aachen University, Germany. Since 2012 he has been a Research Scientist with the Chair of Multimedia Communications and Signal Processing, Friedrich-Alexander University Erlangen-N\"urnberg (FAU), Germany, where he received his Dr.-Ing. degree in 2017.

In 2018 and 2019, he worked as a PostDoc-Fellow at \'Ecole de technologie sup\'erieure in collaboration with Summit Tech Multimedia, Montr\'eal, Canada on energy efficient VR technologies. Since 2019, he is with Friedrich-Alexander University Erlangen-N\"urnberg as a senior scientist. His current research interests include energy efficient video communications and video coding. 
\end{IEEEbiography}

\begin{IEEEbiography}[{\includegraphics[width=1in,height=1.25in,clip,keepaspectratio]{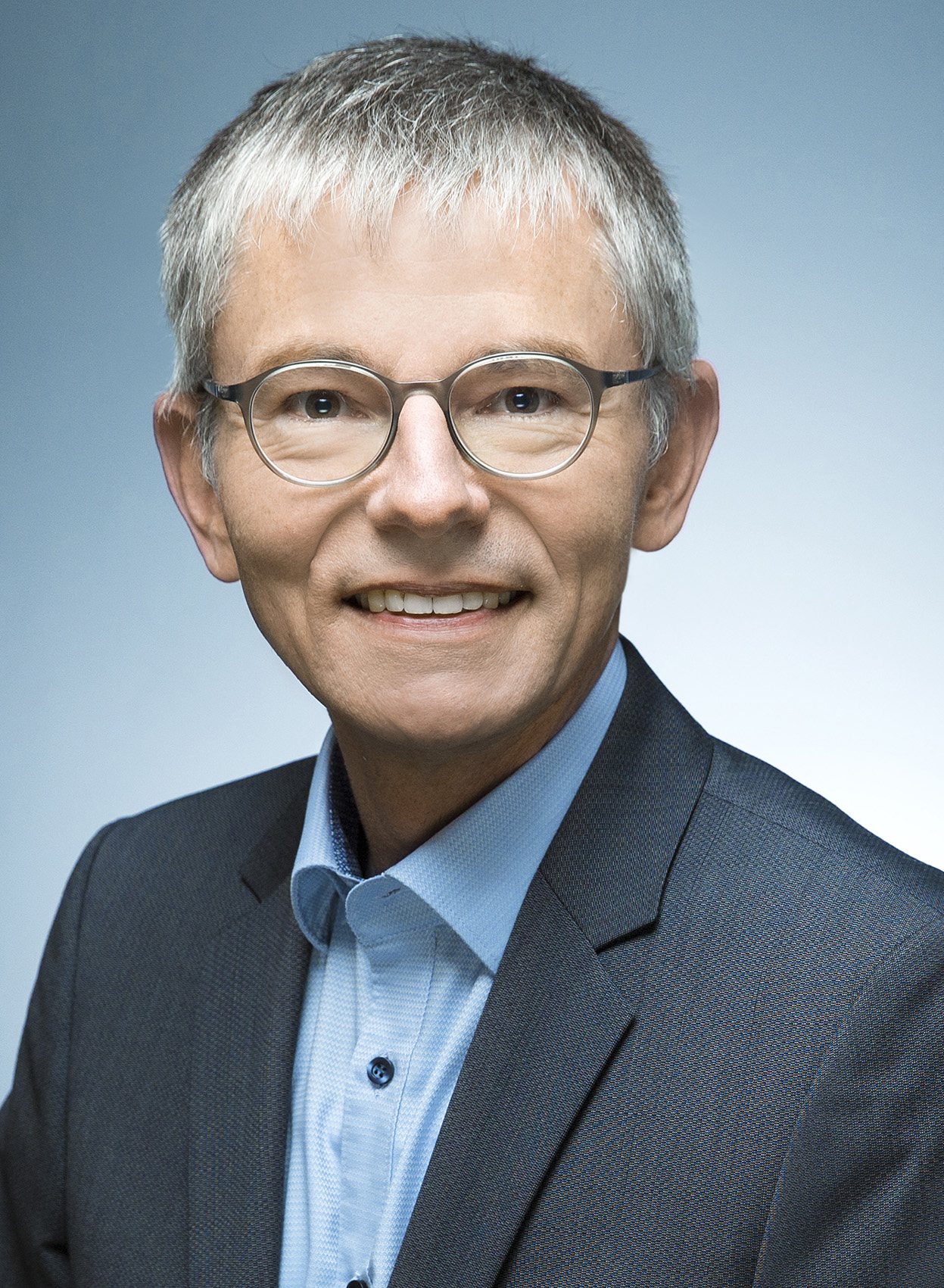}}]{Andr\'e Kaup} (Fellow, IEEE) received the Dipl.-Ing. and Dr.-Ing. degrees in electrical engineering from RWTH Aachen University, Aachen, Germany, in 1989 and 1995, respectively.
 
He joined Siemens Corporate Technology, Munich, Germany, in 1995, and became the Head of the Mobile Applications and Services Group in 1999. Since 2001, he has been a Full Professor and the Head of the Chair of Multimedia Communications and Signal Processing at Friedrich-Alexander University Erlangen-Nuremberg (FAU), Germany. From 2005 to 2007 he was Vice Speaker of the DFG Collaborative Research Center 603. From 2015 to 2017, he served as the Head of the Department of Electrical Engineering and the Vice Dean of the Faculty of Engineering at FAU. He has authored around 400 journal and conference papers and has over 120 patents granted or pending. His research interests include image and video signal processing and coding, and multimedia communication.
 
Dr. Kaup is a member of the IEEE Multimedia Signal Processing Technical Committee and the Scientific Advisory Board of the German VDE/ITG. In 2018, he was elected as a Full Member with the Bavarian Academy of Sciences. He was a Siemens Inventor of the Year 1998 and received the 1999 ITG Award and several IEEE Best Paper Awards. His group won the Grand Video Compression Challenge from the Picture Coding Symposium 2013. The Faculty of Engineering with FAU and the State of Bavaria honored him with Teaching Awards, in 2015 and 2020, respectively. He served as an Associate Editor for IEEE Transactions on Circuits and Systems for Video Technology. He was a Guest Editor for IEEE Journal of Selected Topics in Signal Processing.
\end{IEEEbiography}

\end{document}